\documentclass[twocolumn,twocolappendix]{aastex701}

\usepackage{bm}
\usepackage{mhchem}
\usepackage{hyperref}
\usepackage{amsmath}
\usepackage{amssymb}
\usepackage{graphicx,epstopdf}
\usepackage{epsfig}
\usepackage{color}
\usepackage{colortbl}
\usepackage{xcolor}
\usepackage{aas_macros}
\definecolor{lightgray}{rgb}{0.9,0.9,0.9}
\definecolor{lightergray}{rgb}{0.95,0.95,0.95}
\definecolor{lightblue}{rgb}{0.9,0.9,1}
\definecolor{lightgreen}{rgb}{0.9,1,0.9}

\def\gtaprx {\lower .1ex\hbox{\rlap{\raise .6ex\hbox{\hskip .3ex
	{\ifmmode{\scriptscriptstyle >}\else
		{$\scriptscriptstyle >$}\fi}}}
	\kern -.4ex{\ifmmode{\scriptscriptstyle \sim}\else
		{$\scriptscriptstyle\sim$}\fi}}}
\def\ltaprx {\lower .1ex\hbox{\rlap{\raise .6ex\hbox{\hskip .3ex
	{\ifmmode{\scriptscriptstyle <}\else
		{$\scriptscriptstyle <$}\fi}}}
	\kern -.4ex{\ifmmode{\scriptscriptstyle \sim}\else
		{$\scriptscriptstyle\sim$}\fi}}}

\newcommand{\cutt}[1]{\textcolor{blue}{}}

\newcommand{\C}{{\ensuremath{^{12}\mathrm{C}}}}

\newcommand{\N}{{\ensuremath{^{14} \mathrm{N}}} }

\PassOptionsToPackage{table}{xcolor}
\usepackage{xcolor}
\usepackage{makecell}
\usepackage{physics}
\usepackage{comment}
\definecolor{smsZzero}{HTML}{D9D9D9}
\definecolor{smsZ5}{HTML}{CCEBC5}
\definecolor{smsZ4}{HTML}{B3CDE3}
\definecolor{smsZ3}{HTML}{FBB4AE}
\definecolor{smsZ2}{HTML}{DECBE4}

\shorttitle{SMSs as LRDs}
\shortauthors{Nandal et al.}

\begin{document}

\title{Pulsational mass loss from supermassive stars creates the compact shells of Little Red Dots}

\author[0000-0003-1927-4397]{Devesh Nandal}
\affiliation{Center for Astrophysics, Harvard and Smithsonian, 60 Garden St, Cambridge, MA 02138, USA}
\email{deveshnandal@yahoo.com}

\author[0000-0002-7924-3253]{Igor V. Chilingarian}
\affiliation{Center for Astrophysics, Harvard and Smithsonian, 60 Garden St, Cambridge, MA 02138, USA}
\email{igor.chilingarian@cfa.harvard.edu}

\author[0000-0003-0685-3525]{Chris Nagele}
\affiliation{Department of Physics and Astronomy, Johns Hopkins University, Baltimore MD 21218, USA}
\email{cnagele1@jh.edu}

\author[0000-0002-0302-2577]{John Chisholm}
\email{chishom@austin.utexas.edu}
\affiliation{Department of Astronomy, The University of Texas at Austin, Austin, TX 78712, USA}
\affiliation{Cosmic Frontier Center, The University of Texas at Austin, Austin, TX 78712, USA} 

\author[0000-0002-8686-8737]{Franz E. Bauer}
\affiliation{Instituto de Alta Investigaci{\'{o}}n, Universidad de Tarapac{\'{a}}, Casilla 7D, Arica, 1010000, Chile}
\email{franz.e.bauer@gmail.com}

\author[0000-0003-4330-287X]{Abraham Loeb}
\affiliation{Center for Astrophysics, Harvard and Smithsonian, 60 Garden St, Cambridge, MA 02138, USA}
\email{aloeb@cfa.harvard.edu}

\correspondingauthor{Devesh Nandal}
\email{deveshnandal@yahoo.com}

\begin{abstract}
Little Red Dots (LRDs) have emerged as one of the central puzzles of the JWST era. Their spectra increasingly require dense gas close to the source, yet the physical origin of that cocoon-like structure remains unclear. We examine whether late pulsational mass loss from supermassive stars (SMS) leads to dense gas cocoons. We analyze five GENEC models at different metallicities with characteristic masses of order $10^5\,M_\odot$,  following them through post-accretion evolution with radial pulsation calculations and general relativistic (GR) stability diagnostics. Mass loss during the final stages of evolution occurs not as a steady wind, but through discrete strange-mode ejection episodes. In the $Z=10^{-2}\,Z_\odot$ model, four late episodes last $41$--$282$ yr and eject $10$--$348\,M_\odot$ each, for a total loss of $(4.8-10)\times10^2\,M_\odot$; the final episode alone contributes $\simeq 73\%$ of that budget. Since the last episode dominates the mass-loss, it is the only event sufficiently massive enough to leave behind a compact, optically thick shell extending out to 0.4 pc that reproduces the LRD dense gas cocoon. Final ejecta are H/He dominated but chemically distinctive, with a robust nitrogen-rich composition, $\log(\mathrm{N/O})\simeq0.13$ and $\log(\mathrm{C/O})\simeq-0.23$. SMS reaches GR instability at an age of $\sim 1$ Myr and collapses in $\sim10^4$ s, retaining $\sim 99\%$ all of its mass. Across the full metallicity range from Pop III to $10^{-2}\,Z_\odot$, this shell-ejection channel persists. SMSs therefore provides a physically motivated origin for the compact cocoon-like structure implied by LRDs, while remaining the natural progenitors of the massive black hole seeds invoked in direct collapse scenario.

\end{abstract}


\section{Introduction}

The existence of $\gtrsim 10^9\,M_\odot$ black holes within the first billion years remains a demanding constraint on models of seed formation in the early universe, and JWST has now placed this problem in a new observational context. Supermassive stars provide one of the clearest routes to heavy seeds because their collapse can occur from progenitors that already reach $\sim 10^5\,M_\odot$, avoiding the need for prolonged super-Eddington growth from light remnants. Early work established the basic structure and stability of radiation-dominated stars and clarified how relativistic instability can drive their collapse \citep{hf63, chandra64, fowler66, af72b, fuller86}. More recent studies showed that rapid accretion can assemble stars of $10^4$--$10^6\,M_\odot$ and revived SMSs as plausible progenitors of heavy black-hole seeds \citep{bl03, begel06, hoy12, sak15, tyr17, hle18b}. Related work has also broadened the environments in which such objects may form, including dense stellar systems and mildly enriched gas \citep{regan19, wise19, Reinoso2023, nan24a, nan25a}.

JWST has given this problem a new observational setting through the discovery of Little Red Dots (LRDs), a population of compact high-redshift sources with red rest-optical continua, broad Balmer emission, and prominent spectral structure around the Balmer break \citep{Furtak2023, Harikane2023, Matthee2024, Kocevski2025, Akins2025, degraaff2025,Setton2025, Naidu2025}.  A growing body of work now points to dense gas close to the source as a central ingredient in shaping both the continuum and the emission-line appearance of at least part of the LRD population \citep{Baggen2024, Inayoshi2025, Ji2025, Rusakov2026, kokorev25, Asada2026, Matthee2026}. This has shifted attention not only to the nature of the central engine, but also to the origin of the compact circumsource material, it's stability, evolution, time-scales, and survivability itself. Any SMS-based interpretation must therefore explain how dense gas can be placed on compact scales and the kinematics at the stage when the source becomes observable as an LRD, rather than treating that material as an external assumption \citep{Inayoshi2025, degraaff2025b, Maiolino2025}. 
 
Recent studies have shown that SMS spectra can reproduce several defining LRD features, while other models invoke SMSs embedded in massive self-gravitating accretion structures \citep{Zwick2025, Nandal2026, Chisholm2026}. However, a key remaining question is whether SMSs can also produce the surrounding gas required by these interpretations. Pulsational mass loss offers one possible channel. The envelopes of luminous radiation-dominated stars are prone to strongly nonadiabatic strange-mode instability, and related pulsations have long been studied in massive stars and supergiants \citep{Kiriakidis1993, Glatzel1994, Saio1998, Godart2011, Saio2013, Sonoi2014, Yadav2018}. In the SMS context, most previous work has focused on whether pulsations limit continued growth during rapid accretion, or on how instability develops as the star approaches relativistic collapse \citep{Inayoshi2013, Nakauchi2020, Saio2024}. Much less attention has been given to whether late pulsational episodes can expel weakly bound envelope material and thereby explain the compact shell with properties relevant to LRD phenomenology. 

In this Letter, we test the following scenario for the Balmer-break/LRD regime. After accretion ends, the SMS contracts, ignites hydrogen burning, and re-expands into a phase of strange-mode instability. A small number of discrete pulsational ejections then remove weakly bound envelope material and build a compact dense cocoon with a characteristic composition around the star. The SMS itself continues evolving toward GR instability and later collapses into a heavy black-hole seed. Section~\ref{sec:methods} describes the stellar models and pulsation framework. Section~\ref{sec:results} presents the resulting mass-loss episodes, shell properties, and ejecta composition. Section~\ref{sec:discussion} discusses the implications for LRDs and heavy-seed formation and section~\ref{sec:conc} summarizes our findings.

\section{Methods}
\label{sec:methods}

\subsection{Stellar models}

We analyse five accreting GENEC SMS models spanning $Z/Z_\odot = 0$, $10^{-5}$, $10^{-4}$, $10^{-3}$, and $10^{-2}$. Each sequence begins from a fully convective $10\,M_\odot$ seed and grows at a near-constant accretion rate of $\dot{M}=1\,M_\odot\,{\rm yr^{-1}}$ to a characteristic mass of order $10^5\,M_\odot$. This controlled accretion history is chosen to isolate the late post-accretion pulsations and their connection to the final fate of the star. In more realistic environments, the accretion rate can be highly variable \citep{Woods2021,Nandal2026b}. If \(\dot{M}\) drops below the critical value for maintaining an inflated supergiant envelope, about \(2.5\times10^{-2}\,M_\odot\,{\rm yr^{-1}}\) during pre-main-sequence evolution, the star can contract blueward toward the ZAMS \citep{nan23}. Such excursions could trigger additional strange-mode episodes before the post-accretion phase and would increase the total pulsation linked mass loss over the full growth history. Our focus here is narrower since the LRD relevant cocoon is set by the final pre collapse ejection, because earlier ejecta have much longer to expand and become optically diluted. The models are then followed through their post-accretion evolution. We use the advanced GENEC nuclear network \citep{Nandal2026b}, which follows species up to the Fe group, and each stored structure contains more than 1200 radial layers.

Our pulsation analysis uses every available stored model along each sequence. Additional details of the input structure, preprocessing, and quality checks are given in Appendix~\ref{app:extended_methods}.

\subsection{Radial pulsations and episodic mass loss}

The envelopes of SMSs are luminous, weakly bound, and strongly radiation dominated \citep{hos12, tyr20a}. Strange modes are envelope-confined pulsations that arise in stars with high \(L/M\) and short thermal times in their outer layers. Unlike \(\epsilon\)-modes, which are tied to nuclear burning, or \(\kappa\)-modes, which rely on opacity driving, strange modes are favored when radiation pressure and rapid radiative diffusion alter the usual phase relation between pressure and density in the outer envelope. SMSs naturally enter this regime, so strongly nonadiabatic surface layers can support strange-mode behaviour \citep{Saio1998,Saio2013,Sonoi2014}. We therefore solve the linear adiabatic radial problem in Newtonian gravity for the lowest few modes of each stored model, and also carry out a complementary general-relativistic radial calculation following \citet{Saio2024}. The Newtonian modes provide the default eigenfunctions for the later analysis, while the GR calculation tracks the approach to GR instability.

We then evaluate driving and damping on these adiabatic eigenfunctions, rather than solving the full complex nonadiabatic problem. This quasi-nonadiabatic step provides a growth proxy. Physically, it asks whether the radiative envelope does net positive work on the oscillation over a cycle, so that the mode gains energy and grows, or instead loses more energy to damping than it receives from driving:
\begin{equation}
E_{\rm mode}=\int \frac{1}{2}\rho \omega_r^2 \xi_r^2\, dV,
\qquad
\gamma=\frac{1}{2E_{\rm mode}}\int \dot p_{\rm tot}\, dV ,
\label{eq:methods_gamma}
\end{equation}
where \(E_{\rm mode}\) is the mode energy, \(\rho\) the local density, \(\omega_r\) the real mode frequency, \(\xi_r\) the radial displacement eigenfunction, and \(\dot p_{\rm tot}\) the local work rate per unit volume from all driving and damping contributions. The quantity \(\gamma\) measures the corresponding net growth rate.

We then map the inferred driving through energy conservation and pulsation-linked mass-loss estimate,
\begin{equation}
\dot M = \frac{2\eta L_{\rm drive}}{v_{\rm esc,eff}^2},
\label{eq:methods_mdot}
\end{equation}
and cap each estimate by the mass accessible above the inferred driving region. We do not interpret these values as steady winds. Instead, we group unstable models into discrete ejection episodes and derive the mass-loss histories, launch conditions, and shell properties used below. Full diagnostics and assumptions are given in Appendix~\ref{app:extended_methods}.

\subsection{GR stability and collapse follow-up}

We assess the final fate of the models with complementary GR stability diagnostics \citep{Saio2024,Haem2021A&A...647A..83H,nag22}, and remap unstable snapshots to a 1D GR hydrodynamics code with a 52-isotope nuclear network and neutrino cooling \citep{nag21}. In the baseline calculation, the unstable SMS proceeds to black-hole formation rather than disruption. Further details of the GR criteria, snapshot selection, and collapse evolution are given in Appendix~\ref{app:extended_methodsA10} and~\ref{app:extended_methodsA11} .

\begin{figure*}
\centering
\includegraphics[width=0.78\textwidth]{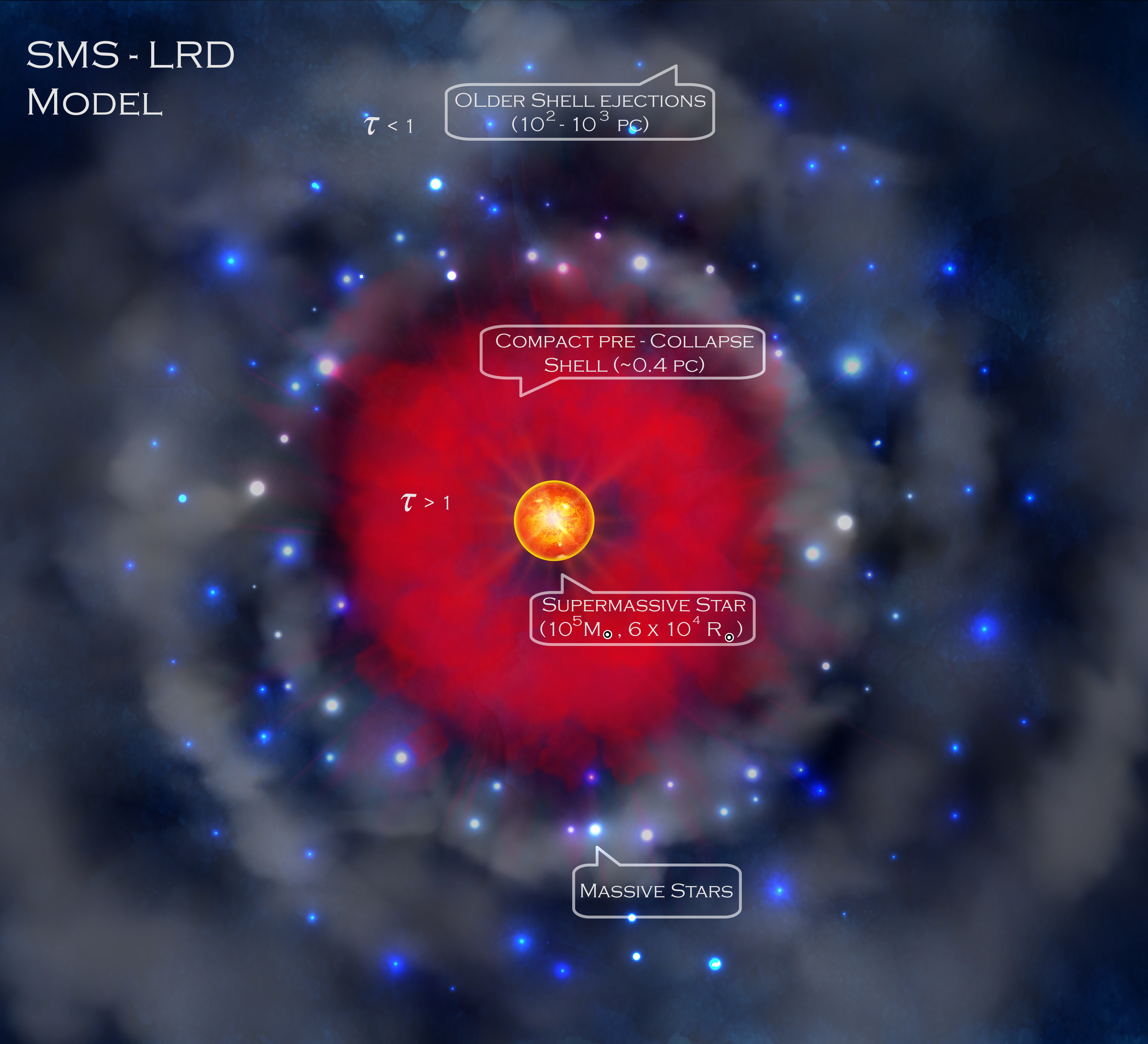}
\caption{Schematic illustration of the SMS pathway explored in this Letter, from the end of accretion to collapse. After accretion ends, the star contracts, ignites hydrogen burning, and re-expands into a late phase of strange-mode instability. Pulsation-driven mass loss then proceeds through discrete ejection episodes that remove weakly bound envelope material. The earlier shells expand to large radii, whereas the final pre-collapse ejection remains compact and dense, setting the immediate circumstellar environment relevant to the LRD phase. The ejecta also carries a characteristic abundance pattern. The SMS then continues toward GR instability and ultimately collapses into a heavy black-hole seed. The drawing is schematic and not to scale.}
\label{fig:concept}
\end{figure*}

\begin{figure*}
\centering
\includegraphics[width=0.49\textwidth]{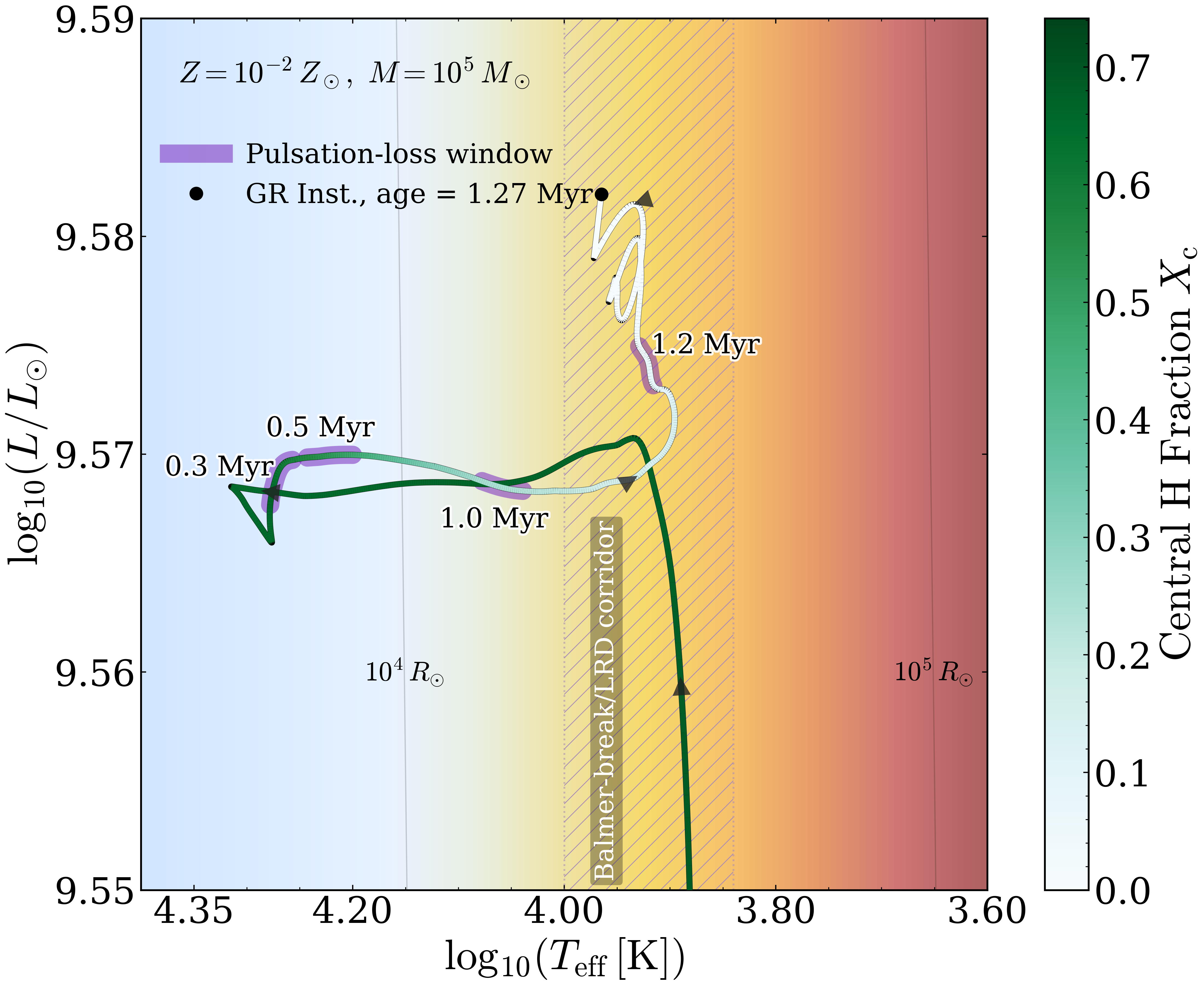}
\hfill
\includegraphics[width=0.49\textwidth]{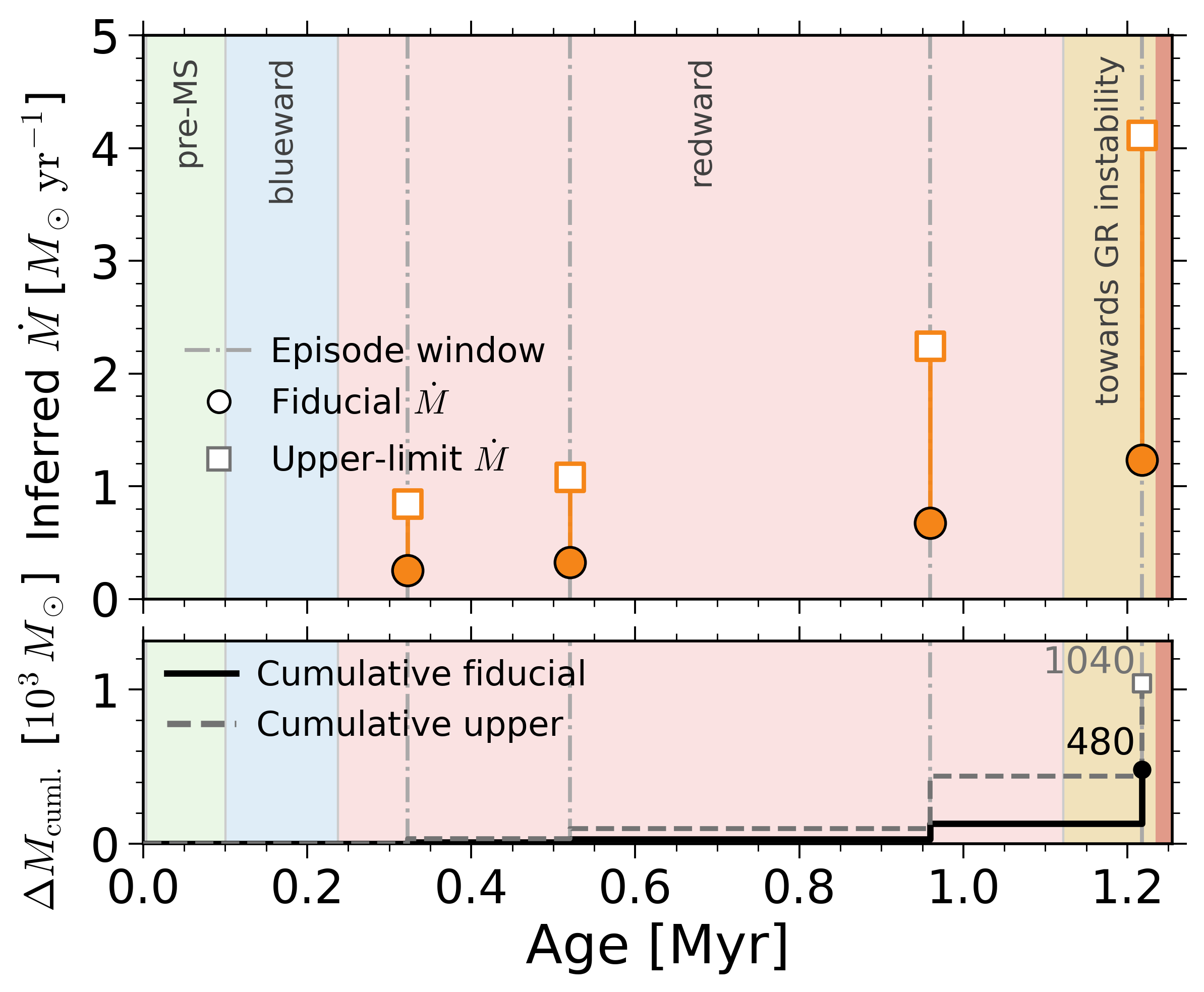}
\caption{
{\it Left:} Post-accretion Hertzsprung--Russell evolution of the $10^{5}\,M_\odot$, $Z=10^{-2}\,Z_\odot$ model commences at $\log(L/L_\odot)=9.57$, $\log T_{\rm eff}=3.90$. The track is colored by the central hydrogen mass fraction, $X_{\rm c}$. Purple segments mark phases with pulsation-driven mass loss, the black circle marks the onset of GR instability, and the hatched band indicates the Balmer-break/LRD corridor. Thin black curves show lines of constant radius.
{\it Right:} Pulsation-driven mass-loss history of the same sequence. The top panel shows the episode-averaged mass-loss rate for the fiducial and upper branches, and the bottom panel shows the cumulative ejected mass. Four discrete episodes occur at $0.322$, $0.520$, $0.959$, and $1.218$ Myr, with both the rates and integrated ejecta increasing toward late times. The final pre-collapse episode dominates the total mass lost.
}
\label{fig:HRD_massloss}
\end{figure*}

\section{Results}
\label{sec:results}
Figure~\ref{fig:concept} summarizes the evolutionary picture that emerges from our analysis. After accretion ends, the SMS contracts, ignites hydrogen burning, and re-expands into a late phase of strange-mode instability. In this phase, pulsation-driven mass loss occurs through a small number of discrete shell ejections. The earlier ejections expand to large radii, while the final pre-collapse ejection remains compact, optically thick, chemically distinctive, and therefore sets the immediate circumstellar environment relevant to the LRD phase. The SMS then continues to GR instability and ultimately collapses into a heavy black-hole seed. In the rest of this section, we quantify this sequence and identify which ejection episode dominates the circumstellar environment at collapse.

Of the five models in our grid, we focus first on the \(Z=10^{-2}\,Z_\odot\) model, which is the closest match to the low but non-zero metallicity regime often associated with LRDs and their nearby analogues. No radiative-wind prescription is imposed, so the mass loss discussed below is entirely pulsational. The accreting phase is pulsationally quiet and is not pursued further here; a fuller account of the pre-main sequence is given in \citet{nan23,Nandal2025b, Nandal2026}.

\subsection{Post-accretion evolution of the \(Z=10^{-2}\,Z_\odot\) model}
\label{sec:results_hrd}
The post-accretion track is shown in the left panel of Fig.~\ref{fig:HRD_massloss}, where color encodes the declining central H mass fraction. Once accretion ends, the inflated envelope is no longer maintained, and the star contracts on a Kelvin--Helmholtz timescale. The track moves blueward from $\log(L/L_\odot)=9.57$ and $\log T_{\rm eff}=3.92$ to a maximum temperature of $\log T_{\rm eff}=4.32$ at nearly unchanged luminosity, $\log(L/L_\odot)=9.55$. This contraction lasts $\simeq 1.35\times10^5$ yr and remains entirely free of pulsation-driven mass loss.

The contraction ends at $X_{\rm c}\simeq0.69$, where the track bends back to the red. By then the energy budget is dominated by CNO burning, and the track migrates to lower $T_{\rm eff}$ again as the outward flux does work in expanding the envelope \citep{nan23,nan24c,Nandal2026}. The star therefore returns to a red-supergiant-like configuration, not on the short thermal timescale of the earlier contraction, but on the much slower nuclear timescale. Over the next $\sim9\times10^5$ yr it evolves to $\log(L/L_\odot)=9.57$, $\log T_{\rm eff}=3.90$, and age $1.1$ Myr, by which point the central H fraction has fallen to $X_{\rm c}\simeq0.10$. During this re-expansion, the inflated radiation-dominated envelope becomes increasingly weakly bound, while the outer layers remain strongly nonadiabatic, favoring strange-mode driving. The first three pulsation-driven mass loss episodes appear along this long redward drift, marked by the purple segments in the left panel of Fig.~\ref{fig:HRD_massloss}.

The final stage is spent at near constant effective temperature highlighted in the same panel. From $1.10$ to $1.27$ Myr, the star remains in this region and undergoes its fourth, and final, pulsation-loss episode. By age $1.27$ Myr, when $X_{\rm c}\simeq0.01$, the model satisfies the GR instability criterion and enters dynamical collapse. Despite this large excursion across the HR diagram, the luminosity stays confined to the narrow range $\log(L/L_\odot)=9.55$--$9.57$ along the full post-accretion track.

\subsection{Pulsation-driven mass-loss history}

The four purple windows in the left panel of Fig.~\ref{fig:HRD_massloss} map directly onto the four discrete ejection episodes in the right panel, centered at 0.322, 0.520, 0.959, and 1.218 Myr. The mass loss is therefore highly selective in time. It is not spread smoothly across the post-accretion evolution, but concentrated into several well-separated outbursts.

The right panel of Fig.~\ref{fig:HRD_massloss} shows that the strength of these events rises systematically with time. On the fiducial branch, the mean mass-loss rates increase from 0.25 and 0.32 $M_\odot\,{\rm yr}^{-1}$ in the first two episodes to 0.67 and 1.23 $M_\odot\,{\rm yr}^{-1}$ in the third and fourth. The upper branch follows the same progression, with rates of 0.84, 1.08, 2.25, and 2.13 $M_\odot\,{\rm yr}^{-1}$. The effective durations of each episode also lengthen from 41.1 and 60.6 yr to 151.2 and 281.9 yr. The late episodes are therefore stronger both in their instantaneous rates and in their finite durations.

The same pattern appears even more clearly in the integrated ejecta masses in the bottom window of the right panel in Fig.~\ref{fig:HRD_massloss}. The fiducial episode masses are 10.4, 19.7, 101.9, and 348.0 $M_\odot$, while the upper branch gives 34.6, 65.6, 339.5, and 599.9 $M_\odot$. Summed over the full sequence, the cumulative mass lost reaches $4.80\times10^2\,M_\odot$ on the fiducial branch and $1.04\times10^3\,M_\odot$ on the upper branch. Even the larger of these remains at only the $\sim1\%$ level of the $\sim10^5\,M_\odot$ star. The pulsations therefore do not disrupt the SMS as a whole. They remove a modest but physically important fraction of the outer envelope. The final episode dominates this budget, contributing $\simeq73\%$ of the cumulative fiducial loss and $\simeq58\%$ of the cumulative upper-limit loss.

The physical trend is equally clear. All four active episodes are classified as strange-mode dominated in our channel decomposition, consistent with strongly nonadiabatic, surface-centered driving in the radiation-dominated envelope. The first two events remain modest. The third is already an order of magnitude stronger in ejected mass than the first, and the fourth becomes the defining mass-loss event of the sequence. This strengthening is not caused by a steadily growing reservoir, since the accessible mass cap stays near $6\times10^2\,M_\odot$ across all four episodes. A more natural interpretation is that, as the star returns to an extended cool configuration and approaches both the Balmer-break corridor and the onset of GR instability, the coupling between the pulsation and the outer envelope becomes markedly more effective. 

\subsection{Event durations and launch conditions}

\begin{figure*}
\center
\begin{tabular}{c}
\includegraphics[width=\hsize]{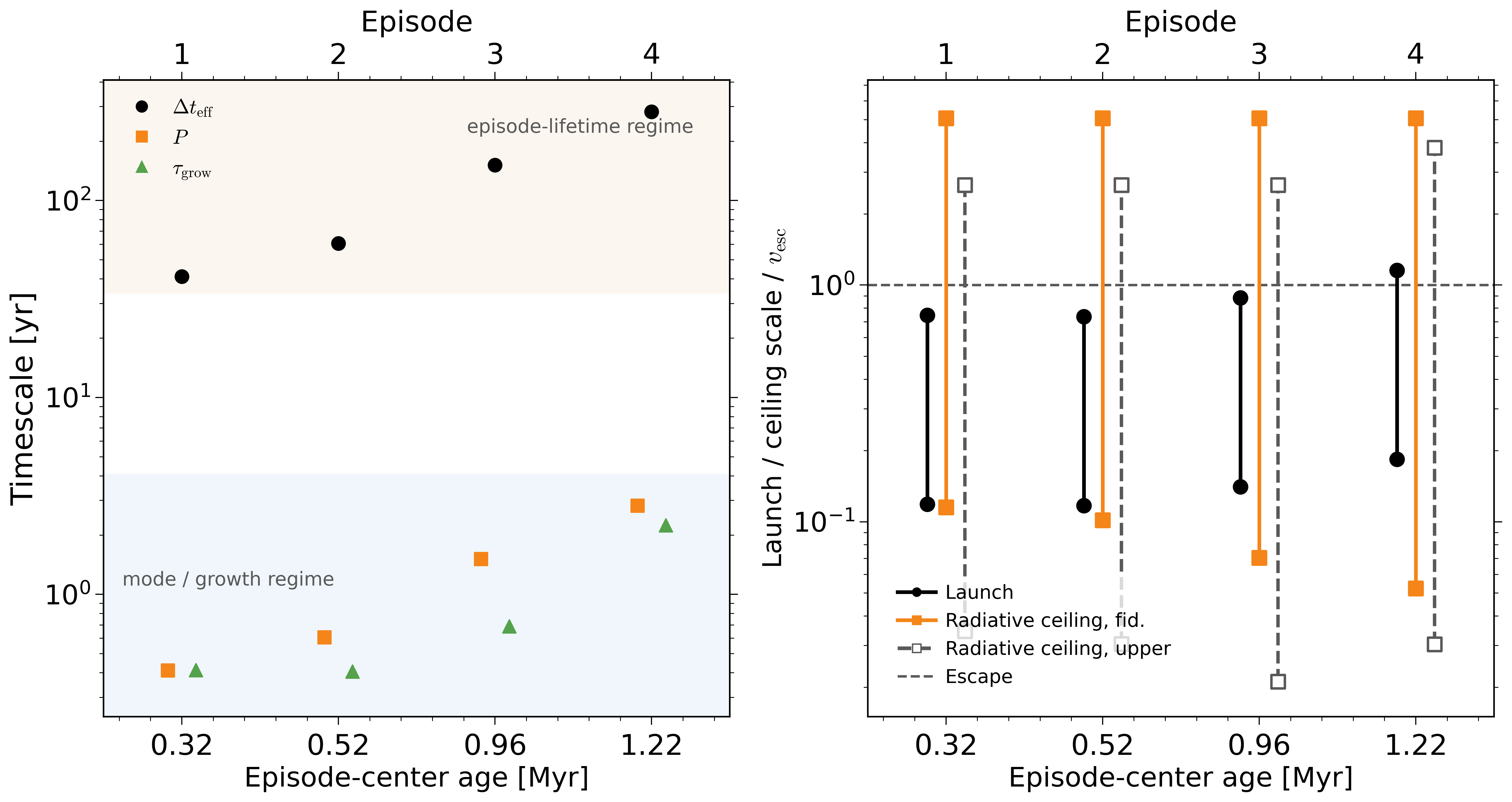}   \\
\end{tabular}
\caption{Timescale and velocity diagnostics for the four pulsation-driven mass-loss episodes in the \(Z=10^{-2}\,Z_\odot\) \(10^5\,M_\odot\) sequence. {\it Left:} effective episode durations compared with the mode period, the linear growth e-folding time, and the local half-gap between adjacent stored models. The events last \(41.1\), \(60.6\), \(151.2\), and \(281.9\) yr, corresponding to \(\sim100\) pulsation cycles in each case, and remain far shorter than the local evolutionary spacing. {\it Right:} structural escape speed, characteristic pulsation speeds \(R/P\) and \(2\pi R/P\), and two radiative reference scales on the fiducial branch: the lower point gives the single-scattering momentum scale \(v_{L/c}\), while the upper point gives the photon-tiring scale \(v_{\rm tir}\).}
\label{fig:timescales_velocity}
\end{figure*}

The two panels of Fig.~\ref{fig:timescales_velocity} place the four mass-loss episodes on complementary physical scales. In the left panel, the effective durations (\(\Delta t_{\rm eff}\)) are 41.1, 60.6, 151.2, and 281.9 yr for episodes 1--4. These values are far longer than the linear pulsation periods (P), 0.411, 0.606, 1.512, and 2.819 yr, and also longer than the corresponding growth e-folding times (\(\tau_{\rm grow}\)), 0.412, 0.406, 0.688, and 2.242 yr. Each event therefore lasts for \(\sim 100P\), or roughly \(10^2\) to \(2\times10^2\) growth times. At the same time, the events remain far shorter than the local evolutionary half-gaps, which are \(9.9\times10^4\), \(8.8\times10^4\), \(7.0\times10^4\), and \(6.0\times10^4\) yr. They are therefore neither single-cycle impulses nor quasi-steady winds, but intrinsically finite pulsation episodes.

The right panel shows how the launch conditions evolve from one episode to the next. The structural escape speed declines from 2587 and 2283 km s\(^{-1}\) in the first two episodes to 1583 and 1176 km s\(^{-1}\) in the third and fourth. Over the same interval, the linear pulsation speed scale \(R/P\) remains much smaller, falling only from 307 to 216 km s\(^{-1}\). Taken on its own, that scale is too small to unbind the outer layers. The sinusoidal scale \(2\pi R/P\), however, rises to a large fraction of the escape speed in the first three events, with \(v_{2\pi R/P}/v_{\rm esc}\simeq 0.74\), 0.73, and 0.88, and reaches 1.15 in the final episode. By the last event, the natural pulsation speed scale has therefore become comparable to, and slightly larger than, the local escape threshold.

The two radiative reference scales in Fig.~\ref{fig:timescales_velocity} separate a momentum constraint from an energy constraint. The lower radiative point is the single-scattering momentum scale, \(v_{L/c}=L_\star/(\dot M c)\). Because these events are discrete pulsation-driven ejections rather than steady winds, we use \(v_{L/c}\) only as a benchmark for whether a purely radiative, momentum-limited outflow could launch the inferred event-averaged mass flux. On the fiducial branch, \(v_{L/c}\) decreases from 298 and 232 km s\(^{-1}\) in the first two episodes to 111 and 61.5 km s\(^{-1}\) in the third and fourth, and remains well below \(v_{\rm esc}\) throughout. The upper radiative point is the photon-tiring scale, which instead remains comfortably above escape in all four cases, ranging from \(1.31\times10^4\) to \(5.96\times10^3\) km s\(^{-1}\). The global radiative energy budget is therefore sufficient, but a purely momentum-limited radiative outflow would still struggle on its own. The limiting factor is how efficiently strange-mode pulsations couple that available energy to the weakly bound outer envelope.

Taken together, Fig.~\ref{fig:timescales_velocity} shows why the final outburst dominates the sequence. It is the longest-lived event, it occurs when the envelope is least tightly bound, and it is the only episode for which \(2\pi R/P\) overtakes \(v_{\rm esc}\). The key result is therefore clear: the luminosity can power the ejection, but only the late strange-mode pulsations couple that energy efficiently enough to launch the outer envelope. As shown in Sec.~\ref{sec:results_shells} and discussed further in Sec.~\ref{shell_prop}, this also makes the final shell the only one likely to remain optically relevant during an LRD-like phase: the earlier ejecta expand and dilute, whereas the last event can still form a compact reprocessing layer on timescales of only a few to a few tens of years.

\subsection{Circumstellar outcome at near the onset of collapse}
\label{sec:results_shells}

\begin{figure}
\center
\begin{tabular}{c}
\includegraphics[width=\columnwidth]{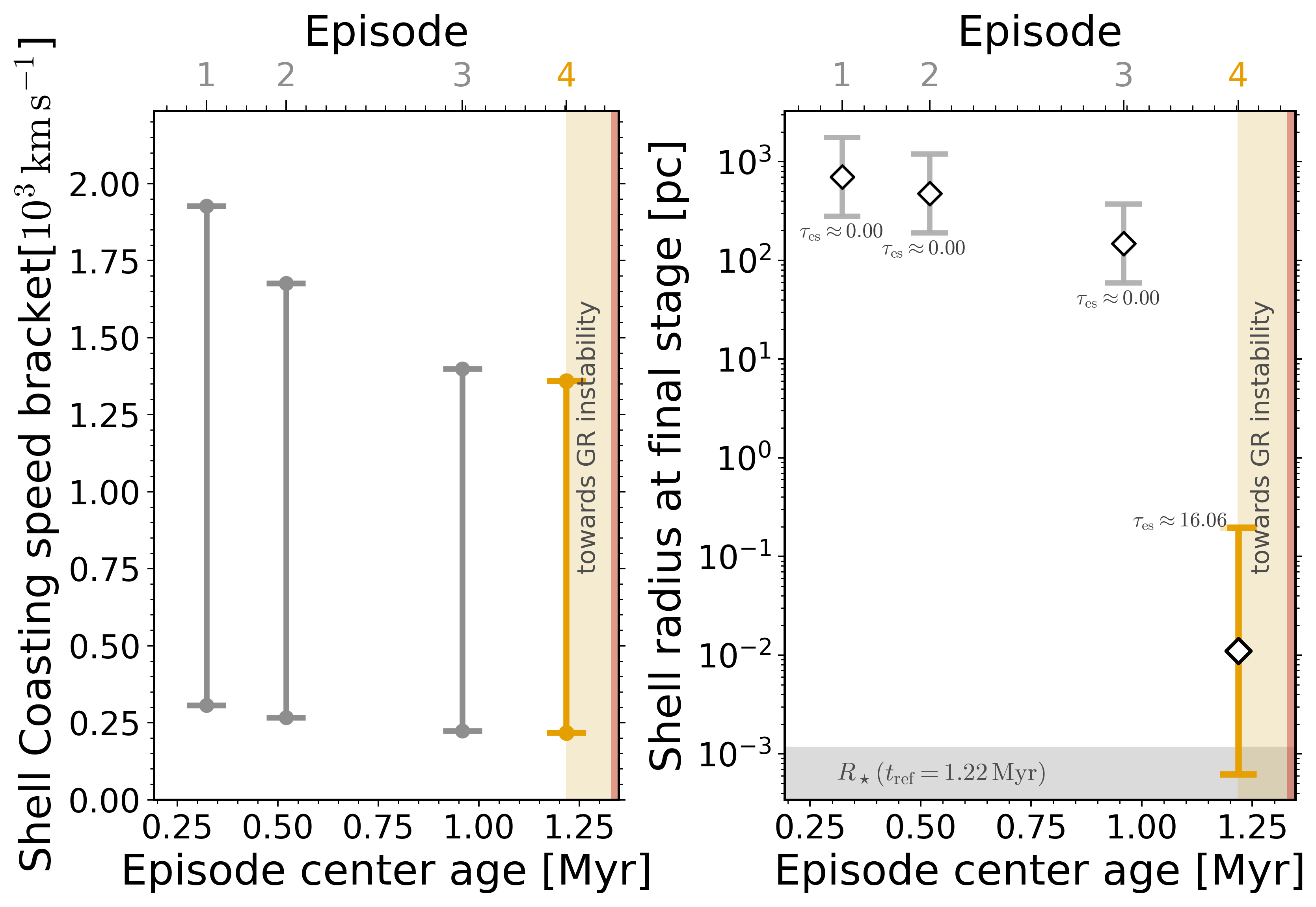}   \\
\end{tabular}
\caption{Shell morphology implied by the four pulsation-driven mass-loss episodes at the end of the \(Z=10^{-2}\,Z_\odot\) \(10^5\,M_\odot\) model. The reference epoch is the end of the fourth episode. The first three ejections have already coasted to large radii and are optically thin, with shell bands spanning \(\sim 59\) to \(1766\) pc and negligible optical depth. The fourth event remains compact, extending from the stellar radius to only \(0.392\) pc, and is the only shell that remains optically important at collapse, with \(\tau_{\rm geo}\simeq 8\). The immediate circumstellar environment is therefore set by the final strange-mode eruption, while the earlier episodes survive only as diffuse outer relics.}
\label{fig:shell_outcome}
\end{figure}

Figure~\ref{fig:shell_outcome} maps the four strange-mode mass-loss episodes into the circumstellar structural parameters present at the end of the model. The launch-speed brackets, inferred from Fig.~\ref{fig:timescales_velocity}, vary only modestly from one episode to the next, with \(v_{\rm low}\simeq 0.22\)--\(0.31\times10^3\) km s\(^{-1}\) and \(v_{\rm high}\simeq 1.36\)--\(1.93\times10^3\) km s\(^{-1}\). The circumstellar outcome is therefore shaped less by a large change in launch speed than by a large difference in coasting time: the first three eruptions have had hundreds of thousands of years to travel outward, whereas the fourth is created only shortly before collapse.

This contrast is clear in the right panel of Fig.~\ref{fig:shell_outcome}. By the onset of collapse, the first three fiducial episodes occupy shell-like bands with inner radii of \(281.0\), \(190.4\), and \(59.0\) pc and outer radii of \(1765.5\), \(1196.3\), and \(370.6\) pc. The fourth event is qualitatively different. Its inner edge remains at the stellar radius, \(6.23\times10^{-4}\) pc, and its outer edge reaches only \(0.392\) pc. The result is a strongly stratified circumstellar structure: three old, highly extended shells surround a single compact inner ejection that remains close to the star.

The optical-depth contrast is equally sharp. For the first three episodes, the geometric optical depths are negligible, \(\tau_{\rm geo}\simeq 1.18\times10^{-10}\), \(4.89\times10^{-10}\), and \(2.63\times10^{-8}\), and even the corresponding inner-edge values remain tiny, at \(7.42\times10^{-10}\), \(3.07\times10^{-9}\), and \(1.66\times10^{-7}\). The fourth shell again stands apart: \(\tau_{\rm geo}\simeq 8.04\), \(\tau_{\rm in}\simeq 5.06\times10^3\), and \(\tau_{\rm out}\simeq 1.28\times10^{-2}\). Within this approximation, only the last event can provide an optically important circumstellar shell at collapse. These values are order-of-magnitude diagnostics based on a constant opacity, \(\kappa=0.34\,{\rm cm^2\,g^{-1}}\), rather than full radiative-transfer solutions. A conceptually similar behavior is observed in R Coronae Borealis-type variable stars, F-to-G type hydrogen-deficient supergiants, which experience episodic fading of their optical brightness by a factor of up-to 5000 explained by the ejection of optically thick dusty shells \citep[see e.g., ][]{Iben1996}.

Finally, the shell masses reinforce the same picture. On the fiducial branch, the first three events eject \(10.4\), \(19.7\), and \(101.9\,M_\odot\), while the final episode contributes \(348.0\,M_\odot\). By collapse, the first three strange-mode eruptions have diluted into very extended, optically negligible relics, whereas the final event remains both compact and massive enough to set the immediate circumstellar environment.

\subsection{Composition of the final ejecta}
\label{sec:results_composition}

\begin{figure*}
\center
\begin{tabular}{c}
\includegraphics[width=\hsize]{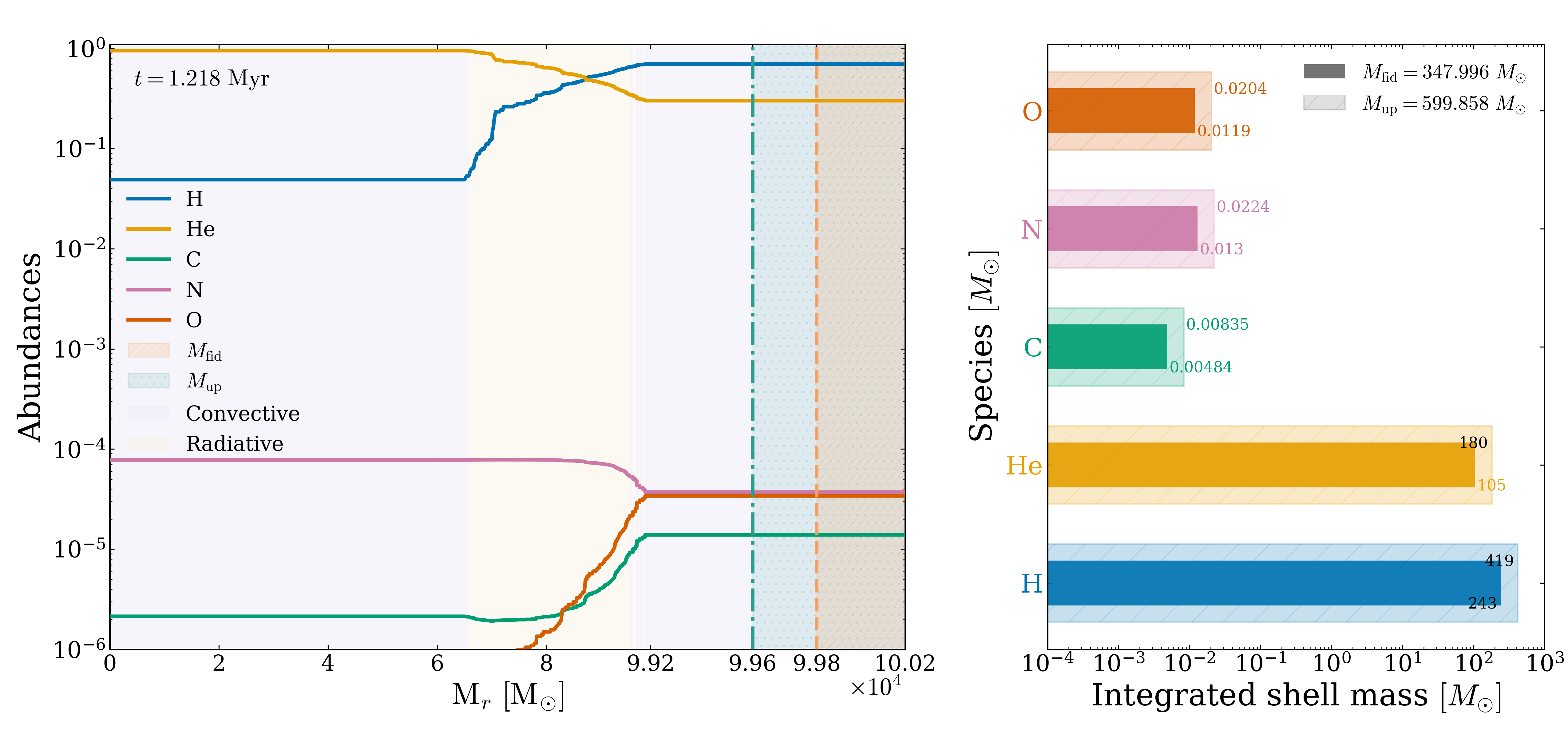}   \\
\end{tabular}
\caption{Composition of the final pulsation-driven shell at \(t=1.218\) Myr. The {\it left panel} shows the abundance profile of the outer envelope together with the fiducial and upper ejection windows; both sample the outer radiative layers, with the upper window reaching slightly deeper. The {\it right panel} shows the integrated shell masses of H, He, C, N, and O for the fiducial and upper shells. In both cases the ejecta are dominated by H and He, but the trace CNO composition is nitrogen rich.}
\label{fig:abundance}
\end{figure*}

Since Fig.~\ref{fig:shell_outcome} shows that the immediate pre-collapse environment is set by the fourth and final eruption, we now examine the composition of that shell in Fig.~\ref{fig:abundance}. The left panel shows the abundance profile at \(t=1.218\) Myr together with the fiducial and upper ejection windows. Both sample the outer radiative envelope and remain overwhelmingly H/He rich. The upper window reaches slightly deeper, but the overall CNO pattern remains similar in the two cases.

The right panel of Fig.~\ref{fig:abundance} shows the integrated shell masses. In the fiducial shell, the ejecta contain \(243\,M_\odot\) of H and \(105\,M_\odot\) of He, together with \(4.84\times10^{-3}\), \(1.3\times10^{-2}\), and \(1.1\times10^{-2}\,M_\odot\) in C, N, and O. The upper shell follows the same pattern, with \(419\,M_\odot\) of H, \(180\,M_\odot\) of He, and \(8.35\times10^{-3}\), \(2.24\times10^{-2}\), and \(2.04\times10^{-2}\,M_\odot\) in C, N, and O. In both cases the shell is therefore H/He dominated by mass, while the trace heavy-element pattern is nitrogen rich, with \(N>O>C\).

The logarithmic number ratios make the same point more clearly. For the fiducial shell we obtain \(\log(N/O)\simeq0.13\) and \(\log(C/O)\simeq-0.23\). For the upper shell the corresponding values are \(\log(N/O)\simeq0.10\) and \(\log(C/O)\simeq-0.26\). The \(\mathrm{He/H}\) number ratio is also nearly unchanged, with \(\mathrm{He/H}\simeq0.108\) and \(0.107\) for the fiducial and upper shells. The key result is that the nitrogen excess is robust. It survives the uncertainty in how deeply the final eruption reaches into the outer envelope.

This is the main message of Fig.~\ref{fig:abundance}. The final strange-mode shell is not merely H rich. It carries a distinctive chemical signature. In our earlier SMS calculations, enhanced N/O emerged as a recurring abundance fingerprint of SMS ejecta \citep{nan24a,nan24b,Nandal2025b,Nandal2025}. The shell shown here therefore provides an explicit composition prediction for the compact circumstellar material present at collapse, with direct observational diagnostics in N/O, C/O, and He/H.

This nitrogen rich composition connects the final shell to a broader abundance signature now emerging in JWST spectroscopy. SMSs have long been discussed as sources of proton-capture products in globular clusters, including the C--N, Na--O, Mg--Al, and Na--F abundance patterns \citep{Denissenkov2014}. More recently, N/O-enhanced high-redshift systems such as GN-z11 have renewed interest in SMS-like enrichment in proto-cluster environments \citep{bun23, cam23}. A related signature is now appearing in some LRDs and LRD-like AGN. The ultraluminous LRD A2744-45924 shows strong N\,{\sc iii}] and N\,{\sc iv}] emission, CANUCS-LRD-z8.6 shows N\,{\sc iv}] in a low-metallicity AGN-like spectrum, and a \(z=6.98\) LRD shows narrow N\,{\sc v} with large inferred N\,{\sc v}/C\,{\sc iv} and N\,{\sc v}/He\,{\sc ii} ratios \citep{Labbe2024,Tripodi2025,Tang2025, Morel2025}. These UV line ratios are not one-to-one abundance measurements, because ionization structure and resonant transfer can also affect them. They nevertheless show that nitrogen-rich dense gas is an important observable in this population. In this context, the final strange mode shell provides a concrete SMS origin composition for the compact cocoon, and links the LRD shell picture to the forming globular-cluster scenario in which enhanced He and N are expected \citep{Chisholm2026}.

Taken together, these results define a late-time SMS mass-loss channel in which rare strange-mode outbursts eject only a modest fraction of the envelope, while the final pre-collapse event leaves behind the compact, chemically distinctive shell most relevant to the LRD phase.

\section{Discussion}
\label{sec:discussion}

\subsection{From SMS to a Direct Collapse Black Hole}

The final fate of the $Z=10^{-2}\,Z_\odot$ SMS is set by general-relativistic instability. We compute three estimates of the onset of the GR radial instability based on individual stellar snapshots. A linear adiabatic GR criterion of \citet{Saio2024} and post-Newtonian estimate of \citet{Haemmerle2021} both place the onset of instability at an age of roughly $1.3\,{\rm Myr}$, while a relativistic adiabatic analysis which accounts for non-linear perturbations finds a slightly earlier onset at $0.9\,{\rm Myr}$ \citep{nag22}. The 1D GR hydrodynamics follow-up confirms that this earlier model collapses to a black hole. All three criteria agree that the SMS will live for approximately one mega-year and will form a black hole after losing a small fraction of its envelope. The GR hydrodynamics calculation shows that this black hole will form within roughly $10^4$ s of the instability, and that the vast majority of the stellar material will fall into the black hole. 
In this sense, the SMS provides a direct pathway to a $\sim10^5\,M_\odot$ direct-collapse black hole.

The models presented in this work are non rotating. A fully consistent rotating post-accretion GENEC calculation for the same \(10^5\,M_\odot\) mass range has not yet been computed, so rotation remains an important uncertainty. Existing rotating GENEC SMS models on the main sequence extend to \(10^3\)--\(10^4\,M_\odot\), and show lifetime extensions of \(7\%\), \(14\%\), and \(20\%\) at \(10^3\,M_\odot\), \(5\times 10^3\,M_\odot\), and \(9\times 10^3\,M_\odot\), respectively \citep{Nandal2025b}. These models also encounter the \(\Omega\Gamma\) limit even for initially slow rotation, \(v/v_{\rm crit}=0.05\)--\(0.10\). For accreting \(10^5\,M_\odot\) SMSs, \citet{Haemmerle2021} found that growth to supermassive scales requires accreted angular momentum of only \(\sim 1\%\) of the Keplerian value, giving \(v_{\rm surf}\lesssim 0.1\)--\(0.2\,v_{\rm crit,1}\) and negligible centrifugal deformation. Rotation can nevertheless affect the final age, angular-momentum content, and GR-instability point. In post treated rotating hylotropic models, \citet{Haemmerle2021} found that \(f=0.3\%\)--\(0.5\%\) can increase the maximum GR-stable mass by an order of magnitude. We therefore expect rotation to shift the timing of collapse and possibly the angular distribution of the ejecta. However, it should not remove the strange mode mass loss channel identified here, because the driving is set mainly by the large \(L/M\), low gas-pressure fraction, and short thermal times of the radiation-dominated envelope \citep{Kiriakidis1993,Glatzel1994,Saio1998,Saio2013,Sonoi2014,Haemmerle2019, saio24}. For the slow surface rotation expected in SMSs, the leading-order radial ejection geometry should remain close to spherical, although differential rotation may modify the covering factor. Rotating post-accretion SMS models and pulsation calculations will be explored in future work.

We caveat that we have not included all relevant forms of mass loss in these calculations, and therefore may be underestimating the amount of mass ejected. In particular, line driven winds \citep{nag23b,Nagele2023MNRAS.523.1629N} and thermonuclear driven pulsations \citep{nag22,nag23a,Nagele2024} can both liberate $100\sim1000 \;M_\odot$ of SMS material. The more material ejected, the lighter the remnant black hole, but also the longer the cocoon remains optically thick and resembles an LRD. For instance if strange mode ejecta remains marginally bound, or is decelerated by the dense natal reservoir
and stalls at the shell scale of $0.392\,{\rm pc}$, that ejecta would return on a fallback timescale of $\sim 1.3\times10^4\,{\rm yr}$. Larger ejecta masses and stall radii would lead to longer fall-back times. 
The importance of this fallback mechanism is to prolong the lifetime of the optically thick cocoon and to provide a new reservoir for accretion:
if the returning gas retains sufficient angular momentum, part of it could circularize into a disc, providing a possible link between SMS collapse and a longer-lived, gas-enshrouded accreting source \citep{begel08,Naidu2025}.

We have not included a steady radiative-wind prescription in the fiducial models. This choice is deliberate, because no steady-wind prescription has been calibrated for \(10^5\,M_\odot\), near-Eddington SMS envelopes. The Vink prescription discussed in \citet{Nandal2026b} was developed for line driven winds of hot O/B stars, while the de Jager relation is an empirical HR-diagram fit based on much lower-mass stars; applying either to SMSs therefore requires an extrapolation by orders of magnitude in mass, luminosity, and Eddington factor \citep{Vink2001,deJager1988}. The estimates discussed in Appendix~A.5 should therefore be read as indicative bounds rather than calibrated wind predictions. If real finite metallicity winds are as strong as the extrapolated cool-supergiant values, they would precondition the envelope and could reduce the reservoir available for later strange mode ejection. Our calculation instead isolates the pulsation linked component that follows directly from the computed stellar structure. The unstable layers are identified from the radial eigenfunctions and shell-by-shell work integral, and the ejected mass is capped by the material above the inferred driving region. The strange mode episodes therefore do not simply peel off an externally prescribed surface layer; they tap an extended, radiation dominated envelope in which the mode does appreciable work. Existing rotating GENEC SMS models also find that \(\Omega\Gamma\)-limited mass loss is episodic and removes only a percent-level fraction of the stellar mass during core-H burning \citep{Nandal2025b}. Thus the present models should be read as isolating the structurally driven pulsational component in the regime where the inflated SMS envelope is retained. Calibrated SMS wind calculations are needed to determine the additional steady wind contribution.

\subsection{Pulsation-driven shells as the origin of cool LRD continua}
\label{shell_prop}
Some recent works have argued that at least part of the LRD continuum arises from cool dense gas with characteristic temperatures of a few $10^3$~K \citep{Liu2025,Kido2025,Wang2026, degraaff2025b}. Our models offer a natural way to produce such values. The low observed temperature does not have to trace the hydrostatic SMS surface. It can instead be set by the last optically thick shell ejected by pulsations.

The shell properties in our $Z=10^{-2}\,Z_\odot$ model are already sufficient for this. The luminosity stays near $\log(L/L_\odot)\simeq 9.57$. Re-emitting that luminosity at $4000$~K requires a photospheric radius of only $\simeq 5.9\times10^2$~AU. For $3000$~K, the required radius is $\simeq 1.05\times10^3$~AU. These radii are only a few times larger than the hydrostatic SMS radius for $T_{\rm eff}\sim7000$--$8000$~K, and they remain far inside the final shell extent of $\sim0.392$~pc.

The onset timescale is also short. Using the launch-speed range of the last ejection, $v\simeq2.2\times10^2$--$1.36\times10^3\,{\rm km\,s^{-1}}$, the shell reaches the $4000$~K radius in only $\sim2$--$13$~yr. It reaches the $3000$~K radius in only $\sim4$--$23$~yr. A cool apparent photosphere can therefore be established within only a few years to a few tens of years after ejection. The inferred shell mass, $\sim3.5\times10^2$--$6\times10^2\,M_\odot$, is large enough that only a modest continuum opacity is required for part of the ejecta to maintain $\tau\sim1$ at these radii. Adiabatic expansion and later dust formation would strengthen the same trend. In this picture, an intrinsically hotter SMS can still produce a continuum with an apparent temperature of only $3000$--$4000$~K.

The shell reprocessing picture also clarifies the expected relation between enshrouded LRDs and unobscured SMSs. In this scenario, an SMS would most naturally be selected as an LRD during the interval in which dense gas provides a cool apparent photosphere or substantial continuum reprocessing. Let \(n_{\rm sh}\) be the number density of SMSs observed through an optically important shell, and let \(n_{\rm bare}\) be the number density of unobscured or weakly obscured SMSs. A simple duty cycle estimate gives \(n_{\rm sh}/n_{\rm bare}\sim f_{\rm cov}t_{\rm sh}/t_{\rm SMS}\), where \(f_{\rm cov}\) is the effective angular covering factor of the shell, \(t_{\rm sh}\) is the time for which the shell remains optically relevant, and \(t_{\rm SMS}=1.27\,{\rm Myr}\) is the lifetime of the fiducial model. The final ejection itself lasts \(\simeq2.8\times10^2\,{\rm yr}\), while marginally bound or stalled ejecta could maintain an optically thick cocoon for up to the fallback time, \(\sim10^4\,{\rm yr}\). This gives \(n_{\rm sh}/n_{\rm bare}\sim2\times10^{-4}f_{\rm cov}\)--\(8\times10^{-3}f_{\rm cov}\). Thus unobscured or weakly obscured SMSs may be more common in lifetime terms, but they need not be classified as LRDs because they would lack the red shell-reprocessed continuum. Finite metallicity winds, rotational mass loss, shell asymmetry, and natal gas can all modify \(f_{\rm cov}\) and \(t_{\rm sh}\), so a precise number-density prediction requires radiation-hydrodynamic shell evolution and selection modelling.

\subsection{Predicted observational signatures} 
The predicted photospheric $T_{\rm eff}\sim7000$~K and dusty shells should give the SMS a yellow-hypergiant-like appearance. This is a rare stellar class in the Milky Way and nearby galaxies, but scaled up in luminosity and with a different abundance pattern. Objects such as \emph{IRC+10420} and \emph{IRAS 17163-3907} (the ``Fried Egg Nebula'') show moderately reddened late-A/early-F spectra with characteristic (broad) emission and absorption lines formed in their complex circumstellar envelopes \citep{Humphreys1997,Humphreys2002,Lagadec2011,Wallstrom2015}. SMSs should generate such features, although the different fiducial layer abundance pattern shown in Fig.~\ref{fig:abundance} can modify various absorption-line ratios. We may expect weaker oxygen, carbon, and magnesium, and stronger nitrogen features. Any iron lines would reflect the composition of the birth gas rather than in-situ production during the hydrogen-burning stage.

Pulsation-driven mass loss produces multiple shell components with characteristic launch speeds of order $10^2$--$10^3$ km~s$^{-1}$. This should produce blueshifted multi-component absorption lines, or asymmetric profiles if the components blend, from atomic species such as Na\,{\sc i} and K\,{\sc i} at sufficiently high column density. At the same time, line-driven mass loss at $10^{-2}\,Z_\odot$ should produce more classical wind signatures, including P~Cygni-like profiles in lines such as H$\alpha$ and metastable He\,{\sc i}. Such broad hydrogen and helium features are already seen in LRDs at both low \citep{2025ApJ...980L..34L,2026MNRAS.545f2235J} and high redshift \citep{Matthee2026}.


\noindent
\section{Conclusion}
\label{sec:conc}
In this Letter, we asked whether late pulsational mass loss from supermassive stars can assemble the compact circumstellar material required in SMS-based interpretations of LRDs. Our results show that it can. Late strange-mode pulsations create a small number of discrete pre-collapse ejection episodes that build a compact inner shell without removing more than a modest fraction of the stellar mass. Across the full metallicity range from Pop III to $10^{-2}\,Z_\odot$, this shell-ejection channel persists.

\begin{itemize}
\item \textbf{Late-time SMS mass loss by pulsations is discrete rather than steady.}
The mass loss does not appear as a persistent wind phase. It is concentrated into a finite number of strange-mode ejection episodes, separated by long intervals of relative quiescence. In the $Z=10^{-2}\,Z_\odot$ model, which provides the clearest LRD analogue, only four fiducial episodes occur. They last 41.1, 60.6, 151.2, and 281.9 yr, and eject 10.4, 19.7, 101.9, and $348.0\,M_\odot$.

\item \textbf{The shell relevant to the LRD phase is set by the last major ejection.}
Earlier ejecta have hundreds of thousands of years to coast outward and become optically negligible. The final pre-collapse event remains compact, extending only from the stellar radius to $\sim0.392$ pc, and is the only shell that remains optically important at collapse, with $\tau_{\rm geo}\simeq8$. The observationally relevant shell is therefore not the accumulated residue of prior mass loss. It is the direct product of the final strange-mode outburst.

\item \textbf{The final shell comes with direct, testable predictions.}
In the fiducial $Z=10^{-2}\,Z_\odot$ case, the final shell contains $\sim243\,M_\odot$ of H and $\sim105\,M_\odot$ of He, together with a robust nitrogen-rich trace composition, $\log(N/O)\simeq0.13$ and $\log(C/O)\simeq-0.23$. The ejecta are launched with characteristic speeds of order $10^2$--$10^3\,{\rm km\,s^{-1}}$. The shell therefore carries quantitative predictions in mass, radius, velocity, and composition that can be confronted directly with LRD observations.

\item \textbf{Pulsations reshape the surroundings without erasing the heavy seed.}
In the $Z=10^{-2}\,Z_\odot$ model, the total ejected mass is only $4.80\times10^2\,M_\odot$ on the fiducial branch and $1.04\times10^3\,M_\odot$ on the upper branch, remaining a small fraction of the stellar mass. The star reaches GR instability at an age of $\sim 1$ Myr, and collapses on a timescale of $10^4$ s. The resulting seed therefore remains close to the stellar mass at instability, of order $10^5\,M_\odot$.
\end{itemize}

These results place SMSs at the center of several ideas that are often discussed separately. In the direct-collapse picture, massive black-hole seeds form directly from collapsing protogalactic gas. In quasi-star models, a black hole grows inside a massive hydrostatic envelope. In recent BH$\star$ interpretations of LRDs, a black hole is embedded in dense gas that shapes the emergent spectrum \citep{bl04,begel08,Naidu2025}. Our results suggest that SMS evolution can provide the bridge between these frameworks: it can first assemble the compact shell required for the observable LRD phase, and then collapse almost intact into the heavy seed itself. In this sense, the SMS links the circumstellar structure, the LRD appearance, and the birth of the massive black-hole seed in one continuous evolutionary pathway.

A logical next step is to turn these shell properties into forward models of continua and line diagnostics, so that pulsational shell ejection from SMSs can be tested directly against the observed spectra of LRDs.


\begin{acknowledgments}
DN was supported by the Swiss National Science Fund (SNSF) Postdoctoral Fellowship, grant number: P500-2235464. DN would like to thank Prof. Hideyuki Saio for his foundational work and for his support during the early stages of this project. FEB acknowledges support from ANID-Chile BASAL CATA FB210003 and FONDECYT Regular 1241005. Artificial intelligence tools were used during manuscript preparation for spell-checking and formatting assistance.
\end{acknowledgments}

\bibliographystyle{aasjournalv7}
\bibliography{refs}

@ARTICLE{Morel2025,
       author = {{Morel}, I. and {Schaerer}, D. and {Marques-Chaves}, R. and {Prantzos}, N. and {Charbonnel}, C. and {Brammer}, G. and {Xiao}, M. and {Dessauges-Zavadsky}, M.},
        title = "{Discovery of new N-emitters over a wide redshift range}",
      journal = {arXiv e-prints},
         year = 2025,
        month = nov,
          eid = {arXiv:2511.20484},
        pages = {arXiv:2511.20484},
          doi = {10.48550/arXiv.2511.20484},
archivePrefix = {arXiv},
       eprint = {2511.20484},
 primaryClass = {astro-ph.GA},
       adsurl = {https://ui.adsabs.harvard.edu/abs/2025arXiv251120484M}
}

@ARTICLE{Haemmerle2019,
       author = {{Haemmerl{\'e}}, L. and {Meynet}, G.},
        title = "{Magnetic braking of supermassive stars through winds}",
      journal = {\aap},
         year = 2019,
        month = mar,
       volume = {623},
          eid = {L7},
        pages = {L7},
          doi = {10.1051/0004-6361/201935087},
archivePrefix = {arXiv},
       eprint = {1903.00020},
 primaryClass = {astro-ph.SR},
       adsurl = {https://ui.adsabs.harvard.edu/abs/2019A&A...623L...7H}
}

@ARTICLE{Tang2025,
       author = {{Tang}, Mengtao and {Stark}, Daniel P. and {Plat}, Ad{\`e}le and {Feltre}, Anna and {Katz}, Harley and {Senchyna}, Peter and {Mason}, Charlotte A. and {Whitler}, Lily and {Chen}, Zuyi and {Topping}, Michael W.},
        title = "{JWST/NIRSpec Observations of High-ionization Emission Lines in Galaxies at High Redshift}",
      journal = {\apj},
         year = 2025,
        month = oct,
       volume = {991},
       number = {2},
          eid = {217},
        pages = {217},
          doi = {10.3847/1538-4357/adfd57},
archivePrefix = {arXiv},
       eprint = {2505.06359},
 primaryClass = {astro-ph.GA},
       adsurl = {https://ui.adsabs.harvard.edu/abs/2025ApJ...991..217T}
}

@ARTICLE{Tripodi2025,
       author = {{Tripodi}, Roberta and {Brada{\v{c}}}, Maru{\v{s}}a and {D'Eugenio}, Francesco and {Martis}, Nicholas and {Rihtar{\v{s}}i{\v{c}}}, Gregor and {Willott}, Chris and {Pentericci}, Laura and {Moreschini}, Bianca and {Markevitch}, Maxim and {Asada}, Yoshihisa and {Calabr{\'o}}, Antonello and {Desprez}, Guillaume and {Felicioni}, Giordano and {Gaspar}, Gaia and {Gonzalez}, Anthony H. and {Harshan}, Anishya and {Ji}, Xihan and {Jude{\v{z}}}, Jon and {Lemaux}, Brian C. and {Marconi}, Alessandro and {Markov}, Vladan and {Merida}, Rosa M. and {Napolitano}, Lorenzo and {Noirot}, Ga{\"e}l and {Parente}, Massimiliano and {Peter}, Annika H.~G. and {Robbins}, Luke and {Robertson}, Andrew and {Sarrouh}, Ghassan T.~E. and {Sawicki}, Marcin},
        title = "{A Deep Dive down the Broad-line Region: Permitted O I, Ca II, and Fe II Emission in an Active Galactic Nucleus Little Red Dot at z = 5.3}",
      journal = {\apjl},
         year = 2025,
        month = nov,
       volume = {994},
       number = {1},
          eid = {L6},
        pages = {L6},
          doi = {10.3847/2041-8213/ae13a9},
archivePrefix = {arXiv},
       eprint = {2507.20684},
 primaryClass = {astro-ph.GA},
       adsurl = {https://ui.adsabs.harvard.edu/abs/2025ApJ...994L...6T}
}

@ARTICLE{Labbe2024,
       author = {{Labbe}, Ivo and {Greene}, Jenny E. and {Matthee}, Jorryt and {Treiber}, Helena and {Kokorev}, Vasily and {Miller}, Tim B. and {Kramarenko}, Ivan and {Setton}, David J. and {Ma}, Yilun and {Goulding}, Andy D. and {Bezanson}, Rachel and {Naidu}, Rohan P. and {Williams}, Christina C. and {Atek}, Hakim and {Brammer}, Gabriel and {Cutler}, Sam E. and {Chemerynska}, Iryna and {Cloonan}, Aidan P. and {Dayal}, Pratika and {de Graaff}, Anna and {Fudamoto}, Yoshinobu and {Fujimoto}, Seiji and {Furtak}, Lukas J. and {Glazebrook}, Karl and {Heintz}, Kasper E. and {Leja}, Joel and {Marchesini}, Danilo and {Nanayakkara}, Themiya and {Nelson}, Erica J. and {Oesch}, Pascal A. and {Pan}, Richard and {Price}, Sedona H. and {Shivaei}, Irene and {Sobral}, David and {Suess}, Katherine A. and {van Dokkum}, Pieter and {Wang}, Bingjie and {Weaver}, John R. and {Whitaker}, Katherine E. and {Zitrin}, Adi},
        title = "{An unambiguous AGN and a Balmer break in an Ultraluminous Little Red Dot at z=4.47 from Ultradeep UNCOVER and All the Little Things Spectroscopy}",
      journal = {arXiv e-prints},
         year = 2024,
        month = dec,
          eid = {arXiv:2412.04557},
        pages = {arXiv:2412.04557},
          doi = {10.48550/arXiv.2412.04557},
archivePrefix = {arXiv},
       eprint = {2412.04557},
 primaryClass = {astro-ph.GA},
       adsurl = {https://ui.adsabs.harvard.edu/abs/2024arXiv241204557L}
}

@ARTICLE{Denissenkov2014,
       author = {{Denissenkov}, P.~A. and {Hartwick}, F.~D.~A.},
        title = "{Supermassive stars as a source of abundance anomalies of proton-capture elements in globular clusters}",
      journal = {\mnras},
         year = 2014,
        month = jan,
       volume = {437},
       number = {1},
        pages = {L21-L25},
          doi = {10.1093/mnrasl/slt133},
archivePrefix = {arXiv},
       eprint = {1305.5975},
 primaryClass = {astro-ph.SR},
       adsurl = {https://ui.adsabs.harvard.edu/abs/2014MNRAS.437L..21D}
}

@ARTICLE{Woods2021,
       author = {{Woods}, Tyrone E. and {Patrick}, Samuel and {Elford}, Jacob S. and {Whalen}, Daniel J. and {Heger}, Alexander},
        title = "{On the Evolution of Supermassive Primordial Stars in Cosmological Flows}",
      journal = {\apj},
         year = 2021,
        month = jul,
       volume = {915},
       number = {2},
          eid = {110},
        pages = {110},
          doi = {10.3847/1538-4357/abfaf9},
archivePrefix = {arXiv},
       eprint = {2102.08963},
 primaryClass = {astro-ph.GA},
       adsurl = {https://ui.adsabs.harvard.edu/abs/2021ApJ...915..110W}
}

@ARTICLE{Nagele2024,
       author = {{Nagele}, Chris and {Umeda}, Hideyuki},
        title = "{Formation of black holes from rapidly accreting supermassive stars is not trivial: Simulations of thermonuclear pulsations and explosions}",
      journal = {\prd},
         year = 2024,
        month = sep,
       volume = {110},
       number = {6},
          eid = {L061301},
        pages = {L061301},
          doi = {10.1103/PhysRevD.110.L061301},
archivePrefix = {arXiv},
       eprint = {2408.08352},
 primaryClass = {astro-ph.HE},
       adsurl = {https://ui.adsabs.harvard.edu/abs/2024PhRvD.110c1301N}
}

@ARTICLE{Iben1996,
       author = {{Iben}, Jr., Icko and {Tutukov}, Alexander V. and {Yungelson}, Lev R.},
        title = "{On the Origin of Hydrogen-deficient Supergiants and Their Relation to R Coronae Borealis Stars and Non-DA White Dwarfs}",
      journal = {\apj},
         year = 1996,
        month = jan,
       volume = {456},
        pages = {750},
          doi = {10.1086/176694},
       adsurl = {https://ui.adsabs.harvard.edu/abs/1996ApJ...456..750I}
}

@ARTICLE{Nandal2026b,
       author = {{Nandal}, Devesh and {Chon}, Sunmyon},
        title = "{Growth of Metal-enriched Supermassive Stars by Accretion and Collisions}",
      journal = {\apj},
         year = 2026,
        month = mar,
       volume = {999},
       number = {1},
          eid = {110},
        pages = {110},
          doi = {10.3847/1538-4357/ae40bb},
archivePrefix = {arXiv},
       eprint = {2511.08516},
 primaryClass = {astro-ph.SR},
       adsurl = {https://ui.adsabs.harvard.edu/abs/2026ApJ...999..110N}
}

@ARTICLE{deJager1988,
       author = {{de Jager}, C. and {Nieuwenhuijzen}, H. and {van der Hucht}, K.~A.},
        title = "{Mass loss rates in the Hertzsprung-Russell diagram.}",
      journal = {\aaps},
         year = 1988,
        month = feb,
       volume = {72},
        pages = {259-289},
       adsurl = {https://ui.adsabs.harvard.edu/abs/1988A&AS...72..259D}
}

@ARTICLE{Vink2001,
       author = {{Vink}, Jorick S. and {de Koter}, A. and {Lamers}, H.~J.~G.~L.~M.},
        title = "{Mass-loss predictions for O and B stars as a function of metallicity}",
      journal = {\aap},
         year = 2001,
        month = apr,
       volume = {369},
        pages = {574-588},
          doi = {10.1051/0004-6361:20010127},
archivePrefix = {arXiv},
       eprint = {astro-ph/0101509},
 primaryClass = {astro-ph},
       adsurl = {https://ui.adsabs.harvard.edu/abs/2001A&A...369..574V}
}

@ARTICLE{Nagele2023MNRAS.523.1629N,
       author = {{Nagele}, Chris and {Umeda}, Hideyuki and {Takahashi}, Koh},
        title = "{Evolution and explosions of metal-enriched supermassive stars: proton rich general relativistic instability supernovae}",
      journal = {\mnras},
         year = 2023,
        month = aug,
       volume = {523},
       number = {2},
        pages = {1629-1640},
          doi = {10.1093/mnras/stad1522},
archivePrefix = {arXiv},
       eprint = {2301.01941},
 primaryClass = {astro-ph.HE},
       adsurl = {https://ui.adsabs.harvard.edu/abs/2023MNRAS.523.1629N}
}

@ARTICLE{kokorev25,
       author = {{Kokorev}, Vasily and {Chisholm}, John and {Naidu}, Rohan P. and {Fujimoto}, Seiji and {Atek}, Hakim and {Brammer}, Gabriel and {Finkelstein}, Steven L. and {Akins}, Hollis B. and {Berg}, Danielle A. and {Furtak}, Lukas J. and {Fei}, Qinyue and {Hsiao}, Tiger Yu-Yang and {Labb{\'e}}, Ivo and {Matthee}, Jorryt and {Mu{\~n}oz}, Julian B. and {Oesch}, Pascal A. and {Pan}, Richard and {Rinaldi}, Pierluigi and {Saldana-Lopez}, Alberto and {Schaerer}, Daniel and {Volonteri}, Marta and {Zitrin}, Adi},
        title = "{The Deepest GLIMPSE of a Dense Gas Cocoon Enshrouding a Little Red Dot}",
      journal = {arXiv e-prints},
         year = 2025,
        month = nov,
          eid = {arXiv:2511.07515},
        pages = {arXiv:2511.07515},
          doi = {10.48550/arXiv.2511.07515},
archivePrefix = {arXiv},
       eprint = {2511.07515},
 primaryClass = {astro-ph.GA},
       adsurl = {https://ui.adsabs.harvard.edu/abs/2025arXiv251107515K}
}

@ARTICLE{Haem2021A&A...647A..83H,
       author = {{Haemmerl{\'e}}, L.},
        title = "{Establishing a reliable determination of the final mass for rapidly accreting supermassive stars}",
      journal = {\aap},
         year = 2021,
        month = mar,
       volume = {647},
          eid = {A83},
        pages = {A83},
          doi = {10.1051/0004-6361/202039686},
archivePrefix = {arXiv},
       eprint = {2010.08229},
 primaryClass = {astro-ph.HE},
       adsurl = {https://ui.adsabs.harvard.edu/abs/2021A&A...647A..83H}
}

@ARTICLE{bl04,
   author = {{Barkana}, R. and {Loeb}, A.},
    title = "{Unusually Large Fluctuations in the Statistics of Galaxy Formation at High Redshift}",
  journal = {\apj},
   eprint = {arXiv:astro-ph/0310338},
     year = 2004,
    month = jul,
   volume = 609,
    pages = {474-481},
      doi = {10.1086/421079},
   adsurl = {http://adsabs.harvard.edu/abs/2004ApJ...609..474B}
}

@ARTICLE{begel08,
   author = {{Begelman}, M.~C. and {Rossi}, E.~M. and {Armitage}, P.~J.},
    title = "{Quasi-stars: accreting black holes inside massive envelopes}",
  journal = {\mnras},
archivePrefix = "arXiv",
   eprint = {0711.4078},
     year = 2008,
    month = jul,
   volume = 387,
    pages = {1649-1659},
      doi = {10.1111/j.1365-2966.2008.13344.x},
   adsurl = {http://adsabs.harvard.edu/abs/2008MNRAS.387.1649B}
}

@ARTICLE{begel06,
   author = {{Begelman}, M.~C. and {Volonteri}, M. and {Rees}, M.~J.},
    title = "{Formation of supermassive black holes by direct collapse in pre-galactic haloes}",
  journal = {\mnras},
   eprint = {arXiv:astro-ph/0602363},
     year = 2006,
    month = jul,
   volume = 370,
    pages = {289-298},
      doi = {10.1111/j.1365-2966.2006.10467.x},
   adsurl = {http://adsabs.harvard.edu/abs/2006MNRAS.370..289B}
}

@ARTICLE{bl03,
   author = {{Bromm}, V. and {Loeb}, A.},
    title = "{Formation of the First Supermassive Black Holes}",
  journal = {\apj},
   eprint = {arXiv:astro-ph/0212400},
     year = 2003,
    month = oct,
   volume = 596,
    pages = {34-46},
      doi = {10.1086/377529},
   adsurl = {http://adsabs.harvard.edu/abs/2003ApJ...596...34B}
}

@ARTICLE{chandra64,
   author = {{Chandrasekhar}, S.},
    title = "{The Dynamical Instability of Gaseous Masses Approaching the Schwarzschild Limit in General Relativity.}",
  journal = {\apj},
     year = 1964,
    month = aug,
   volume = 140,
    pages = {417},
      doi = {10.1086/147938},
   adsurl = {http://adsabs.harvard.edu/abs/1964ApJ...140..417C}
}

@ARTICLE{fowler66,
   author = {{Fowler}, W.~A.},
    title = "{The Stability of Supermassive Stars}",
  journal = {\apj},
     year = 1966,
    month = apr,
   volume = 144,
    pages = {180},
      doi = {10.1086/148594},
   adsurl = {http://adsabs.harvard.edu/abs/1966ApJ...144..180F}
}

@ARTICLE{fuller86,
   author = {{Fuller}, G.~M. and {Woosley}, S.~E. and {Weaver}, T.~A.},
    title = "{The evolution of radiation-dominated stars. I - Nonrotating supermassive stars}",
  journal = {\apj},
     year = 1986,
    month = aug,
   volume = 307,
    pages = {675-686},
      doi = {10.1086/164452},
   adsurl = {http://adsabs.harvard.edu/abs/1986ApJ...307..675F}
}

@ARTICLE{hf63,
   author = {{Hoyle}, F. and {Fowler}, W.~A.},
    title = "{On the nature of strong radio sources}",
  journal = {\mnras},
     year = 1963,
   volume = 125,
    pages = {169},
   adsurl = {http://adsabs.harvard.edu/abs/1963MNRAS.125..169H}
}

@ARTICLE{hos12,
   author = {{Hosokawa}, T. and {Yoshida}, N. and {Omukai}, K. and {Yorke}, H.~W.
	},
    title = "{Protostellar Feedback and Final Mass of the Second-generation Primordial Stars}",
  journal = {\apjl},
archivePrefix = "arXiv",
   eprint = {1210.3035},
 primaryClass = "astro-ph.CO",
     year = 2012,
    month = dec,
   volume = 760,
      eid = {L37},
    pages = {L37},
      doi = {10.1088/2041-8205/760/2/L37},
   adsurl = {http://adsabs.harvard.edu/abs/2012ApJ...760L..37H}
}

@ARTICLE{nag21,
       author = {{Nagele}, Chris and {Umeda}, Hideyuki and {Takahashi}, Koh and {Yoshida}, Takashi and {Sumiyoshi}, Kohsuke},
        title = "{Neutrino emission from the collapse of {}10$^{4}$ M$_{{\ensuremath{\odot}}}$ Population III supermassive stars}",
      journal = {\mnras},
         year = 2021,
        month = nov,
       volume = {508},
       number = {1},
        pages = {828-841},
          doi = {10.1093/mnras/stab2592},
archivePrefix = {arXiv},
       eprint = {2107.01761},
 primaryClass = {astro-ph.HE},
       adsurl = {https://ui.adsabs.harvard.edu/abs/2021MNRAS.508..828N}
}

@ARTICLE{af72b,
   author = {{Appenzeller}, I. and {Fricke}, K.},
    title = "{Hydrodynamic Model Calculations for Supermassive Stars. II. The Collapse and Explosion of a Nonrotating 5.2 X 10\~{} M0 Star}",
  journal = {\aap},
     year = 1972,
    month = nov,
   volume = 21,
    pages = {285},
   adsurl = {http://adsabs.harvard.edu/abs/1972A%26A....21..285A}
}

@ARTICLE{hoy12,
   author = {{Hosokawa}, T. and {Omukai}, K. and {Yorke}, H.~W.},
    title = "{Rapidly Accreting Supergiant Protostars: Embryos of Supermassive Black Holes?}",
  journal = {\apj},
archivePrefix = "arXiv",
   eprint = {1203.2613},
     year = 2012,
    month = sep,
   volume = 756,
      eid = {93},
    pages = {93},
      doi = {10.1088/0004-637X/756/1/93},
   adsurl = {http://adsabs.harvard.edu/abs/2012ApJ...756...93H}
}

@ARTICLE{sak15,
   author = {{Sakurai}, Y. and {Hosokawa}, T. and {Yoshida}, N. and {Yorke}, H.~W.
	},
    title = "{Formation of primordial supermassive stars by burst accretion}",
  journal = {\mnras},
archivePrefix = "arXiv",
   eprint = {1505.03954},
 primaryClass = "astro-ph.SR",
     year = 2015,
    month = sep,
   volume = 452,
    pages = {755-764},
      doi = {10.1093/mnras/stv1346},
   adsurl = {http://adsabs.harvard.edu/abs/2015MNRAS.452..755S}
}

@ARTICLE{tyr17,
   author = {{Woods}, T.~E. and {Heger}, A. and {Whalen}, D.~J. and {Haemmerl{\'e}}, L. and 
	{Klessen}, R.~S.},
    title = "{On the Maximum Mass of Accreting Primordial Supermassive Stars}",
  journal = {\apjl},
archivePrefix = "arXiv",
   eprint = {1703.07480},
 primaryClass = "astro-ph.SR",
     year = 2017,
    month = jun,
   volume = 842,
      eid = {L6},
    pages = {L6},
      doi = {10.3847/2041-8213/aa7412},
   adsurl = {http://adsabs.harvard.edu/abs/2017ApJ...842L...6W}
}

@ARTICLE{hle18b,
   author = {{Haemmerl{\'e}}, L. and {Woods}, T.~E. and {Klessen}, R.~S. and 
	{Heger}, A. and {Whalen}, D.~J.},
    title = "{The evolution of supermassive Population III stars}",
  journal = {\mnras},
archivePrefix = "arXiv",
   eprint = {1705.09301},
 primaryClass = "astro-ph.SR",
     year = 2018,
    month = feb,
   volume = 474,
    pages = {2757-2773},
      doi = {10.1093/mnras/stx2919},
   adsurl = {http://adsabs.harvard.edu/abs/2018MNRAS.474.2757H}
}

@ARTICLE{wise19,
   author = {{Wise}, J.~H. and {Regan}, J.~A. and {O'Shea}, B.~W. and {Norman}, M.~L. and 
	{Downes}, T.~P. and {Xu}, H.},
    title = "{Formation of massive black holes in rapidly growing pre-galactic gas clouds}",
  journal = {\nat},
archivePrefix = "arXiv",
   eprint = {1901.07563},
     year = 2019,
    month = jan,
   volume = 566,
    pages = {85-88},
      doi = {10.1038/s41586-019-0873-4},
   adsurl = {http://adsabs.harvard.edu/abs/2019Natur.566...85W}
}

@ARTICLE{regan19,
       author = {{Regan}, John A. and {Wise}, John H. and {O'Shea}, Brian W. and
         {Norman}, Michael L.},
        title = "{The emergence of the first star-free atomic cooling haloes in the Universe}",
      journal = {\mnras},
         year = 2020,
        month = feb,
       volume = {492},
       number = {2},
        pages = {3021-3031},
          doi = {10.1093/mnras/staa035},
archivePrefix = {arXiv},
       eprint = {1908.02823},
 primaryClass = {astro-ph.GA},
       adsurl = {https://ui.adsabs.harvard.edu/abs/2020MNRAS.492.3021R}
}

@ARTICLE{tyr20a,
       author = {{Woods}, Tyrone E. and {Heger}, Alexander and {Haemmerl{\'e}}, Lionel},
        title = "{On monolithic supermassive stars}",
      journal = {\mnras},
         year = 2020,
        month = mar,
       volume = {494},
       number = {2},
        pages = {2236-2243},
          doi = {10.1093/mnras/staa763},
archivePrefix = {arXiv},
       eprint = {2003.10467},
 primaryClass = {astro-ph.GA},
       adsurl = {https://ui.adsabs.harvard.edu/abs/2020MNRAS.494.2236W}
}

@ARTICLE{nag20,
       author = {{Nagele}, Chris and {Umeda}, Hideyuki and {Takahashi}, Koh and {Yoshida}, Takashi and {Sumiyoshi}, Kohsuke},
        title = "{The final fate of supermassive M {\ensuremath{\sim}} 5 {\texttimes} {}10$^{4}$ M$_{{\ensuremath{\odot}}}$ Pop III stars: explosion or collapse?}",
      journal = {\mnras},
         year = 2020,
        month = aug,
       volume = {496},
       number = {2},
        pages = {1224-1231},
          doi = {10.1093/mnras/staa1636},
archivePrefix = {arXiv},
       eprint = {2006.08834},
 primaryClass = {astro-ph.HE},
       adsurl = {https://ui.adsabs.harvard.edu/abs/2020MNRAS.496.1224N}
}

@ARTICLE{nag22,
       author = {{Nagele}, Chris and {Umeda}, Hideyuki and {Takahashi}, Koh and {Yoshida}, Takashi and {Sumiyoshi}, Kohsuke},
        title = "{Stability analysis of supermassive primordial stars: a new mass range for general relativistic instability supernovae}",
      journal = {\mnras},
         year = 2022,
        month = dec,
       volume = {517},
       number = {2},
        pages = {1584-1600},
          doi = {10.1093/mnras/stac2495},
archivePrefix = {arXiv},
       eprint = {2205.10493},
 primaryClass = {astro-ph.SR},
       adsurl = {https://ui.adsabs.harvard.edu/abs/2022MNRAS.517.1584N}
}

@ARTICLE{nag23a,
       author = {{Nagele}, Chris and {Umeda}, Hideyuki and {Takahashi}, Koh and {Maeda}, Keiichi},
        title = "{Pulsations of primordial supermassive stars induced by a general relativistic instability; visible to JWST at z > 12}",
      journal = {\mnras},
         year = 2023,
        month = mar,
       volume = {520},
       number = {1},
        pages = {L72-L77},
          doi = {10.1093/mnrasl/slad009},
archivePrefix = {arXiv},
       eprint = {2210.08662},
 primaryClass = {astro-ph.HE},
       adsurl = {https://ui.adsabs.harvard.edu/abs/2023MNRAS.520L..72N}
}

@ARTICLE{nan23,
       author = {{Nandal}, Devesh and {Regan}, John A. and {Woods}, Tyrone E. and {Farrell}, Eoin and {Ekstr{\"o}m}, Sylvia and {Meynet}, Georges},
        title = "{Critical accretion rates for rapidly growing massive Population III stars}",
      journal = {\aap},
         year = 2023,
        month = sep,
       volume = {677},
          eid = {A155},
        pages = {A155},
          doi = {10.1051/0004-6361/202346938},
archivePrefix = {arXiv},
       eprint = {2306.17223},
 primaryClass = {astro-ph.SR},
       adsurl = {https://ui.adsabs.harvard.edu/abs/2023A&A...677A.155N}
}

@ARTICLE{nan24a,
       author = {{Nandal}, D. and {Farrell}, E. and {Buldgen}, G. and {Meynet}, G. and {Ekstr{\"o}m}, S.},
        title = "{The evolution and impact of {\ensuremath{\sim}}3000 M$_{{\ensuremath{\odot}}}$ stars in the early Universe}",
      journal = {\aap},
         year = 2024,
        month = may,
       volume = {685},
          eid = {A159},
        pages = {A159},
          doi = {10.1051/0004-6361/202345997},
archivePrefix = {arXiv},
       eprint = {2309.04435},
 primaryClass = {astro-ph.SR},
       adsurl = {https://ui.adsabs.harvard.edu/abs/2024A&A...685A.159N}
}

@ARTICLE{cam23,
       author = {{Cameron}, Alex J. and {Katz}, Harley and {Rey}, Martin P. and {Saxena}, Aayush},
        title = "{Nitrogen enhancements 440 Myr after the big bang: supersolar N/O, a tidal disruption event, or a dense stellar cluster in GN-z11?}",
      journal = {\mnras},
         year = 2023,
        month = aug,
       volume = {523},
       number = {3},
        pages = {3516-3525},
          doi = {10.1093/mnras/stad1579},
archivePrefix = {arXiv},
       eprint = {2302.10142},
 primaryClass = {astro-ph.GA},
       adsurl = {https://ui.adsabs.harvard.edu/abs/2023MNRAS.523.3516C}
}

@ARTICLE{bun23,
       author = {{Bunker}, Andrew J. and {Saxena}, Aayush and {Cameron}, Alex J. and {Willott}, Chris J. and {Curtis-Lake}, Emma and {Jakobsen}, Peter and {Carniani}, Stefano and {Smit}, Renske and {Maiolino}, Roberto and {Witstok}, Joris and {Curti}, Mirko and {D'Eugenio}, Francesco and {Jones}, Gareth C. and {Ferruit}, Pierre and {Arribas}, Santiago and {Charlot}, Stephane and {Chevallard}, Jacopo and {Giardino}, Giovanna and {de Graaff}, Anna and {Looser}, Tobias J. and {L{\"u}tzgendorf}, Nora and {Maseda}, Michael V. and {Rawle}, Tim and {Rix}, Hans-Walter and {Del Pino}, Bruno Rodr{\'\i}guez and {Alberts}, Stacey and {Egami}, Eiichi and {Eisenstein}, Daniel J. and {Endsley}, Ryan and {Hainline}, Kevin and {Hausen}, Ryan and {Johnson}, Benjamin D. and {Rieke}, George and {Rieke}, Marcia and {Robertson}, Brant E. and {Shivaei}, Irene and {Stark}, Daniel P. and {Sun}, Fengwu and {Tacchella}, Sandro and {Tang}, Mengtao and {Williams}, Christina C. and {Willmer}, Christopher N.~A. and {Baker}, William M. and {Baum}, Stefi and {Bhatawdekar}, Rachana and {Bowler}, Rebecca and {Boyett}, Kristan and {Chen}, Zuyi and {Circosta}, Chiara and {Helton}, Jakob M. and {Ji}, Zhiyuan and {Kumari}, Nimisha and {Lyu}, Jianwei and {Nelson}, Erica and {Parlanti}, Eleonora and {Perna}, Michele and {Sandles}, Lester and {Scholtz}, Jan and {Suess}, Katherine A. and {Topping}, Michael W. and {{\"U}bler}, Hannah and {Wallace}, Imaan E.~B. and {Whitler}, Lily},
        title = "{JADES NIRSpec Spectroscopy of GN-z11: Lyman-{\ensuremath{\alpha}} emission and possible enhanced nitrogen abundance in a z = 10.60 luminous galaxy}",
      journal = {\aap},
         year = 2023,
        month = sep,
       volume = {677},
          eid = {A88},
        pages = {A88},
          doi = {10.1051/0004-6361/202346159},
archivePrefix = {arXiv},
       eprint = {2302.07256},
 primaryClass = {astro-ph.GA},
       adsurl = {https://ui.adsabs.harvard.edu/abs/2023A&A...677A..88B}
}

@ARTICLE{nan24b,
       author = {{Nandal}, Devesh and {Sibony}, Yves and {Tsiatsiou}, Sophie},
        title = "{Fast-rotating massive Population III stars as possible sources of extreme N enrichment in high-redshift galaxies}",
      journal = {\aap},
         year = 2024,
        month = aug,
       volume = {688},
          eid = {A142},
        pages = {A142},
          doi = {10.1051/0004-6361/202348866},
archivePrefix = {arXiv},
       eprint = {2405.11235},
 primaryClass = {astro-ph.GA},
       adsurl = {https://ui.adsabs.harvard.edu/abs/2024A&A...688A.142N}
}

@ARTICLE{nan24c,
       author = {{Nandal}, Devesh and {Regan}, John A. and {Woods}, Tyrone E. and {Farrell}, Eoin and {Ekstr{\"o}m}, Sylvia and {Meynet}, Georges},
        title = "{Explaining the high nitrogen abundances observed in high-z galaxies via population III stars of a few thousand solar masses}",
      journal = {\aap},
         year = 2024,
        month = mar,
       volume = {683},
          eid = {A156},
        pages = {A156},
          doi = {10.1051/0004-6361/202348035},
archivePrefix = {arXiv},
       eprint = {2402.03428},
 primaryClass = {astro-ph.GA},
       adsurl = {https://ui.adsabs.harvard.edu/abs/2024A&A...683A.156N}
}

@ARTICLE{nag23b,
       author = {{Nagele}, Chris and {Umeda}, Hideyuki},
        title = "{Multiple Channels for Nitrogen Pollution by Metal-enriched Supermassive Stars and Implications for GN-z11}",
      journal = {\apjl},
         year = 2023,
        month = may,
       volume = {949},
       number = {1},
          eid = {L16},
        pages = {L16},
          doi = {10.3847/2041-8213/acd550},
archivePrefix = {arXiv},
       eprint = {2304.05013},
 primaryClass = {astro-ph.GA},
       adsurl = {https://ui.adsabs.harvard.edu/abs/2023ApJ...949L..16N}
}

@ARTICLE{saio24,
       author = {{Saio}, Hideyuki and {Nandal}, Devesh and {Ekstr{\"o}m}, Sylvia and {Meynet}, George},
        title = "{Linear adiabatic analysis for general-relativistic instability in primordial accreting supermassive stars}",
      journal = {\aap},
         year = 2024,
        month = sep,
       volume = {689},
          eid = {A169},
        pages = {A169},
          doi = {10.1051/0004-6361/202449971},
archivePrefix = {arXiv},
       eprint = {2406.18040},
 primaryClass = {astro-ph.HE},
       adsurl = {https://ui.adsabs.harvard.edu/abs/2024A&A...689A.169S}
}

@ARTICLE{Maeder2000,
       author = {{Maeder}, A. and {Meynet}, G.},
        title = "{Stellar evolution with rotation. VI. The Eddington and Omega -limits, the rotational mass loss for OB and LBV stars}",
      journal = {\aap},
         year = 2000,
        month = sep,
       volume = {361},
        pages = {159-166},
          doi = {10.48550/arXiv.astro-ph/0006405},
archivePrefix = {arXiv},
       eprint = {astro-ph/0006405},
 primaryClass = {astro-ph},
       adsurl = {https://ui.adsabs.harvard.edu/abs/2000A&A...361..159M}
}

@ARTICLE{nan25a,
       author = {{Nandal}, Devesh and {Buldgen}, Ga{\"e}l and {Whalen}, Daniel J. and {Regan}, John and {Woods}, Tyrone E. and {Tan}, Jonathan C.},
        title = "{Rotating Supermassive Pop III Stars On The Main Sequence}",
      journal = {arXiv e-prints},
         year = 2025,
        month = jun,
          eid = {arXiv:2506.08268},
        pages = {arXiv:2506.08268},
          doi = {10.48550/arXiv.2506.08268},
archivePrefix = {arXiv},
       eprint = {2506.08268},
 primaryClass = {astro-ph.SR},
       adsurl = {https://ui.adsabs.harvard.edu/abs/2025arXiv250608268N}
}

@ARTICLE{Reinoso2023,
       author = {{Reinoso}, Basti{\'a}n and {Klessen}, Ralf S. and {Schleicher}, Dominik and {Glover}, Simon C.~O. and {Solar}, P.},
        title = "{Formation of supermassive stars in the first star clusters}",
      journal = {\mnras},
         year = 2023,
        month = may,
       volume = {521},
       number = {3},
        pages = {3553-3569},
          doi = {10.1093/mnras/stad790},
archivePrefix = {arXiv},
       eprint = {2303.07827},
 primaryClass = {astro-ph.GA},
       adsurl = {https://ui.adsabs.harvard.edu/abs/2023MNRAS.521.3553R}
}

@ARTICLE{Furtak2023,
       author = {{Furtak}, Lukas J. and {Zitrin}, Adi and {Plat}, Ad{\`e}le and {Fujimoto}, Seiji and {Wang}, Bingjie and {Nelson}, Erica J. and {Labb{\'e}}, Ivo and {Bezanson}, Rachel and {Brammer}, Gabriel B. and {van Dokkum}, Pieter and {Endsley}, Ryan and {Glazebrook}, Karl and {Greene}, Jenny E. and {Leja}, Joel and {Price}, Sedona H. and {Smit}, Renske and {Stark}, Daniel P. and {Weaver}, John R. and {Whitaker}, Katherine E. and {Atek}, Hakim and {Chevallard}, Jacopo and {Curtis-Lake}, Emma and {Dayal}, Pratika and {Feltre}, Anna and {Franx}, Marijn and {Fudamoto}, Yoshinobu and {Marchesini}, Danilo and {Mowla}, Lamiya A. and {Pan}, Richard and {Suess}, Katherine A. and {Vidal-Garc{\'\i}a}, Alba and {Williams}, Christina C.},
        title = "{JWST UNCOVER: Extremely Red and Compact Object at z $_{phot}$ ≃ 7.6 Triply Imaged by A2744}",
      journal = {\apj},
         year = 2023,
        month = aug,
       volume = {952},
       number = {2},
          eid = {142},
        pages = {142},
          doi = {10.3847/1538-4357/acdc9d},
archivePrefix = {arXiv},
       eprint = {2212.10531},
 primaryClass = {astro-ph.GA},
       adsurl = {https://ui.adsabs.harvard.edu/abs/2023ApJ...952..142F}}

@ARTICLE{Harikane2023,
       author = {{Harikane}, Yuichi and {Zhang}, Yechi and {Nakajima}, Kimihiko and {Ouchi}, Masami and {Isobe}, Yuki and {Ono}, Yoshiaki and {Hatano}, Shun and {Xu}, Yi and {Umeda}, Hiroya},
        title = "{A JWST/NIRSpec First Census of Broad-line AGNs at z = 4-7: Detection of 10 Faint AGNs with M $_{BH}$ {}10$^{6}$-{}10$^{8}$ M $_{☉}$ and Their Host Galaxy Properties}",
      journal = {\apj},
         year = 2023,
        month = dec,
       volume = {959},
       number = {1},
          eid = {39},
        pages = {39},
          doi = {10.3847/1538-4357/ad029e},
archivePrefix = {arXiv},
       eprint = {2303.11946},
 primaryClass = {astro-ph.GA},
       adsurl = {https://ui.adsabs.harvard.edu/abs/2023ApJ...959...39H}
}

@ARTICLE{Matthee2024,
       author = {{Matthee}, Jorryt and {Naidu}, Rohan P. and {Brammer}, Gabriel and {Chisholm}, John and {Eilers}, Anna-Christina and {Goulding}, Andy and {Greene}, Jenny and {Kashino}, Daichi and {Labbe}, Ivo and {Lilly}, Simon J. and {Mackenzie}, Ruari and {Oesch}, Pascal A. and {Weibel}, Andrea and {Wuyts}, Stijn and {Xiao}, Mengyuan and {Bordoloi}, Rongmon and {Bouwens}, Rychard and {van Dokkum}, Pieter and {Illingworth}, Garth and {Kramarenko}, Ivan and {Maseda}, Michael V. and {Mason}, Charlotte and {Meyer}, Romain A. and {Nelson}, Erica J. and {Reddy}, Naveen A. and {Shivaei}, Irene and {Simcoe}, Robert A. and {Yue}, Minghao},
        title = "{Little Red Dots: An Abundant Population of Faint Active Galactic Nuclei at z {\ensuremath{\sim}} 5 Revealed by the EIGER and FRESCO JWST Surveys}",
      journal = {\apj},
         year = 2024,
        month = mar,
       volume = {963},
       number = {2},
          eid = {129},
        pages = {129},
          doi = {10.3847/1538-4357/ad2345},
archivePrefix = {arXiv},
       eprint = {2306.05448},
 primaryClass = {astro-ph.GA},
       adsurl = {https://ui.adsabs.harvard.edu/abs/2024ApJ...963..129M}
}

@ARTICLE{Kocevski2025,
       author = {{Kocevski}, Dale D. and {Finkelstein}, Steven L. and {Barro}, Guillermo and {Taylor}, Anthony J. and {Calabr{\`o}}, Antonello and {Laloux}, Brivael and {Buchner}, Johannes and {Trump}, Jonathan R. and {Leung}, Gene C.~K. and {Yang}, Guang and {Dickinson}, Mark and {P{\'e}rez-Gonz{\'a}lez}, Pablo G. and {Pacucci}, Fabio and {Inayoshi}, Kohei and {Somerville}, Rachel S. and {McGrath}, Elizabeth J. and {Akins}, Hollis B. and {Bagley}, Micaela B. and {Bowler}, Rebecca A.~A. and {Bisigello}, Laura and {Carnall}, Adam and {Casey}, Caitlin M. and {Cheng}, Yingjie and {Cleri}, Nikko J. and {Costantin}, Luca and {Cullen}, Fergus and {Davis}, Kelcey and {Donnan}, Callum T. and {Dunlop}, James S. and {Ellis}, Richard S. and {Ferguson}, Henry C. and {Fujimoto}, Seiji and {Fontana}, Adriano and {Giavalisco}, Mauro and {Grazian}, Andrea and {Grogin}, Norman A. and {Hathi}, Nimish P. and {Hirschmann}, Michaela and {Huertas-Company}, Marc and {Holwerda}, Benne W. and {Illingworth}, Garth and {Juneau}, St{\'e}phanie and {Kartaltepe}, Jeyhan S. and {Koekemoer}, Anton M. and {Li}, Wenxiu and {Lucas}, Ray A. and {Magee}, Dan and {Mason}, Charlotte and {McLeod}, Derek J. and {McLure}, Ross J. and {Napolitano}, Lorenzo and {Papovich}, Casey and {Pirzkal}, Nor and {Rodighiero}, Giulia and {Santini}, Paola and {Wilkins}, Stephen M. and {Yung}, L.~Y. Aaron},
        title = "{The Rise of Faint, Red Active Galactic Nuclei at z > 4: A Sample of Little Red Dots in the JWST Extragalactic Legacy Fields}",
      journal = {\apj},
         year = 2025,
        month = jun,
       volume = {986},
       number = {2},
          eid = {126},
        pages = {126},
          doi = {10.3847/1538-4357/adbc7d},
archivePrefix = {arXiv},
       eprint = {2404.03576},
 primaryClass = {astro-ph.GA},
       adsurl = {https://ui.adsabs.harvard.edu/abs/2025ApJ...986..126K}
}

@ARTICLE{Akins2025,
       author = {{Akins}, Hollis B. and {Casey}, Caitlin M. and {Lambrides}, Erini and {Allen}, Natalie and {Andika}, Irham T. and {Brinch}, Malte and {Champagne}, Jaclyn B. and {Cooper}, Olivia and {Ding}, Xuheng and {Drakos}, Nicole E. and {Faisst}, Andreas and {Finkelstein}, Steven L. and {Franco}, Maximilien and {Fujimoto}, Seiji and {Gentile}, Fabrizio and {Gillman}, Steven and {Gozaliasl}, Ghassem and {Harish}, Santosh and {Hayward}, Christopher C. and {Hirschmann}, Michaela and {Ilbert}, Olivier and {Kartaltepe}, Jeyhan S. and {Kocevski}, Dale D. and {Koekemoer}, Anton M. and {Kokorev}, Vasily and {Liu}, Daizhong and {Long}, Arianna S. and {McCracken}, Henry Joy and {McKinney}, Jed and {Onoue}, Masafusa and {Paquereau}, Louise and {Renzini}, Alvio and {Rhodes}, Jason and {Robertson}, Brant E. and {Shuntov}, Marko and {Silverman}, John D. and {Tanaka}, Takumi S. and {Toft}, Sune and {Trakhtenbrot}, Benny and {Valentino}, Francesco and {Zavala}, Jorge},
        title = "{COSMOS-Web: The Overabundance and Physical Nature of ``Little Red Dots''{\textemdash}Implications for Early Galaxy and SMBH Assembly}",
      journal = {\apj},
         year = 2025,
        month = sep,
       volume = {991},
       number = {1},
          eid = {37},
        pages = {37},
          doi = {10.3847/1538-4357/ade984},
archivePrefix = {arXiv},
       eprint = {2406.10341},
 primaryClass = {astro-ph.GA},
       adsurl = {https://ui.adsabs.harvard.edu/abs/2025ApJ...991...37A}
}

@ARTICLE{degraaff2025,
       author = {{de Graaff}, Anna and {Rix}, Hans-Walter and {Naidu}, Rohan P. and {Labb{\'e}}, Ivo and {Wang}, Bingjie and {Leja}, Joel and {Matthee}, Jorryt and {Katz}, Harley and {Greene}, Jenny E. and {Hviding}, Raphael E. and {Baggen}, Josephine and {Bezanson}, Rachel and {Boogaard}, Leindert A. and {Brammer}, Gabriel and {Dayal}, Pratika and {van Dokkum}, Pieter and {Goulding}, Andy D. and {Hirschmann}, Michaela and {Maseda}, Michael V. and {McConachie}, Ian and {Miller}, Tim B. and {Nelson}, Erica and {Oesch}, Pascal A. and {Setton}, David J. and {Shivaei}, Irene and {Weibel}, Andrea and {Whitaker}, Katherine E. and {Williams}, Christina C.},
        title = "{A remarkable ruby: Absorption in dense gas, rather than evolved stars, drives the extreme Balmer break of a little red dot at z = 3.5}",
      journal = {\aap},
         year = 2025,
        month = sep,
       volume = {701},
          eid = {A168},
        pages = {A168},
          doi = {10.1051/0004-6361/202554681},
archivePrefix = {arXiv},
       eprint = {2503.16600},
 primaryClass = {astro-ph.GA},
       adsurl = {https://ui.adsabs.harvard.edu/abs/2025A&A...701A.168D}
}

@ARTICLE{Setton2025,
       author = {{Setton}, David J. and {Greene}, Jenny E. and {de Graaff}, Anna and {Ma}, Yilun and {Leja}, Joel and {Matthee}, Jorryt and {Bezanson}, Rachel and {Boogaard}, Leindert A. and {Cleri}, Nikko J. and {Katz}, Harley and {Labbe}, Ivo and {Maseda}, Michael V. and {McConachie}, Ian and {Miller}, Tim B. and {Price}, Sedona H. and {Suess}, Katherine A. and {van Dokkum}, Pieter and {Wang}, Bingjie and {Weibel}, Andrea and {Whitaker}, Katherine E. and {Williams}, Christina C.},
        title = "{Little Red Dots at an Inflection Point: Ubiquitous V-shaped Turnover Consistently Occurs at the Balmer Limit}",
      journal = {\apj},
         year = 2025,
        month = dec,
       volume = {995},
       number = {1},
          eid = {118},
        pages = {118},
          doi = {10.3847/1538-4357/ae1500},
archivePrefix = {arXiv},
       eprint = {2411.03424},
 primaryClass = {astro-ph.GA},
       adsurl = {https://ui.adsabs.harvard.edu/abs/2025ApJ...995..118S}
}

@ARTICLE{Naidu2025,
       author = {{Naidu}, Rohan P. and {Matthee}, Jorryt and {Katz}, Harley and {de Graaff}, Anna and {Oesch}, Pascal and {Smith}, Aaron and {Greene}, Jenny E. and {Brammer}, Gabriel and {Weibel}, Andrea and {Hviding}, Raphael and {Chisholm}, John and {Labb\textbackslash'e}, Ivo and {Simcoe}, Robert A. and {Witten}, Callum and {Atek}, Hakim and {Baggen}, Josephine F.~W. and {Belli}, Sirio and {Bezanson}, Rachel and {Boogaard}, Leindert A. and {Bose}, Sownak and {Covelo-Paz}, Alba and {Dayal}, Pratika and {Fudamoto}, Yoshinobu and {Furtak}, Lukas J. and {Giovinazzo}, Emma and {Goulding}, Andy and {Gronke}, Max and {Heintz}, Kasper E. and {Hirschmann}, Michaela and {Illingworth}, Garth and {Inoue}, Akio K. and {Johnson}, Benjamin D. and {Leja}, Joel and {Leonova}, Ecaterina and {McConachie}, Ian and {Maseda}, Michael V. and {Natarajan}, Priyamvada and {Nelson}, Erica and {Setton}, David J. and {Shivaei}, Irene and {Sobral}, David and {Stefanon}, Mauro and {Tacchella}, Sandro and {Toft}, Sune and {Torralba}, Alberto and {van Dokkum}, Pieter and {van der Wel}, Arjen and {Volonteri}, Marta and {Walter}, Fabian and {Wang}, Bingjie and {Watson}, Darach},
        title = "{A ``Black Hole Star'' Reveals the Remarkable Gas-Enshrouded Hearts of the Little Red Dots}",
      journal = {arXiv e-prints},
         year = 2025,
        month = mar,
          eid = {arXiv:2503.16596},
        pages = {arXiv:2503.16596},
          doi = {10.48550/arXiv.2503.16596},
archivePrefix = {arXiv},
       eprint = {2503.16596},
 primaryClass = {astro-ph.GA},
       adsurl = {https://ui.adsabs.harvard.edu/abs/2025arXiv250316596N}
}

@ARTICLE{Baggen2024,
       author = {{Baggen}, Josephine F.~W. and {van Dokkum}, Pieter and {Brammer}, Gabriel and {de Graaff}, Anna and {Franx}, Marijn and {Greene}, Jenny and {Labb{\'e}}, Ivo and {Leja}, Joel and {Maseda}, Michael V. and {Nelson}, Erica J. and {Rix}, Hans-Walter and {Wang}, Bingjie and {Weibel}, Andrea},
        title = "{The Small Sizes and High Implied Densities of ``Little Red Dots'' with Balmer Breaks Could Explain Their Broad Emission Lines without an Active Galactic Nucleus}",
      journal = {\apjl},
         year = 2024,
        month = dec,
       volume = {977},
       number = {1},
          eid = {L13},
        pages = {L13},
          doi = {10.3847/2041-8213/ad90b8},
archivePrefix = {arXiv},
       eprint = {2408.07745},
 primaryClass = {astro-ph.GA},
       adsurl = {https://ui.adsabs.harvard.edu/abs/2024ApJ...977L..13B}
}

@ARTICLE{Inayoshi2025,
       author = {{Inayoshi}, Kohei and {Maiolino}, Roberto},
        title = "{Extremely Dense Gas around Little Red Dots and High-redshift Active Galactic Nuclei: A Nonstellar Origin of the Balmer Break and Absorption Features}",
      journal = {\apjl},
         year = 2025,
        month = feb,
       volume = {980},
       number = {2},
          eid = {L27},
        pages = {L27},
          doi = {10.3847/2041-8213/adaebd},
archivePrefix = {arXiv},
       eprint = {2409.07805},
 primaryClass = {astro-ph.GA},
       adsurl = {https://ui.adsabs.harvard.edu/abs/2025ApJ...980L..27I}
}

@ARTICLE{Ji2025,
       author = {{Ji}, Xihan and {Maiolino}, Roberto and {{\"U}bler}, Hannah and {Scholtz}, Jan and {D'Eugenio}, Francesco and {Sun}, Fengwu and {Perna}, Michele and {Turner}, Hannah and {Carniani}, Stefano and {Arribas}, Santiago and {Bennett}, Jake S. and {Bunker}, Andrew and {Charlot}, St{\'e}phane and {Cresci}, Giovanni and {Curti}, Mirko and {Egami}, Eiichi and {Fabian}, Andy and {Inayoshi}, Kohei and {Isobe}, Yuki and {Jones}, Gareth and {Juod{\v{z}}balis}, Ignas and {Kumari}, Nimisha and {Lyu}, Jianwei and {Mazzolari}, Giovanni and {Parlanti}, Eleonora and {Robertson}, Brant and {Rodr{\'\i}guez Del Pino}, Bruno and {Schneider}, Raffaella and {Sijacki}, Debora and {Tacchella}, Sandro and {Trinca}, Alessandro and {Valiante}, Rosa and {Venturi}, Giacomo and {Volonteri}, Marta and {Willott}, Chris and {Witten}, Callum and {Witstok}, Joris},
        title = "{BlackTHUNDER ─ A non-stellar Balmer break in a black hole-dominated little red dot at z = 7.04}",
      journal = {\mnras},
         year = 2025,
        month = dec,
       volume = {544},
       number = {4},
        pages = {3900-3935},
          doi = {10.1093/mnras/staf1867},
archivePrefix = {arXiv},
       eprint = {2501.13082},
 primaryClass = {astro-ph.GA},
       adsurl = {https://ui.adsabs.harvard.edu/abs/2025MNRAS.544.3900J}
}

@ARTICLE{Rusakov2026,
       author = {{Rusakov}, V. and {Watson}, D. and {Nikopoulos}, G.~P. and {Brammer}, G. and {Gottumukkala}, R. and {Harvey}, T. and {Heintz}, K.~E. and {Damgaard}, R. and {Sim}, S.~A. and {Sneppen}, A. and {Vijayan}, A.~P. and {Adams}, N. and {Austin}, D. and {Conselice}, C.~J. and {Goolsby}, C.~M. and {Toft}, S. and {Witstok}, J.},
        title = "{Little red dots as young supermassive black holes in dense ionized cocoons}",
      journal = {\nat},
         year = 2026,
        month = jan,
       volume = {649},
       number = {8097},
        pages = {574-579},
          doi = {10.1038/s41586-025-09900-4},
archivePrefix = {arXiv},
       eprint = {2503.16595},
 primaryClass = {astro-ph.GA},
       adsurl = {https://ui.adsabs.harvard.edu/abs/2026Natur.649..574R}
}

@ARTICLE{Asada2026,
       author = {{Asada}, Yoshihisa and {Inayoshi}, Kohei and {Fei}, Qinyue and {Fujimoto}, Seiji and {Willott}, Chris},
        title = "{Origins of the UV continuum and Balmer emission lines in Little Red Dots: observational validation of dense gas envelope models enshrouding the AGN}",
      journal = {arXiv e-prints},
         year = 2026,
        month = jan,
          eid = {arXiv:2601.10573},
        pages = {arXiv:2601.10573},
          doi = {10.48550/arXiv.2601.10573},
archivePrefix = {arXiv},
       eprint = {2601.10573},
 primaryClass = {astro-ph.GA},
       adsurl = {https://ui.adsabs.harvard.edu/abs/2026arXiv260110573A}
}

@ARTICLE{Matthee2026,
       author = {{Matthee}, Jorryt and {Torralba}, Alberto and {Pezzulli}, Gabriele and {Naidu}, Rohan P. and {Chisholm}, John and {Mascia}, Sara and {Greene}, Jenny E. and {Ishikawa}, Yuzo and {Gronke}, Max and {Wuyts}, Stijn and {Bordoloi}, Rongmon and {Brammer}, Gabriel and {Chang}, Seok-Jun and {Eilers}, Anna-Christina and {de Graaff}, Anna and {Hviding}, Raphael E. and {Iani}, Edoardo and {Illingworth}, Garth and {Kashino}, Daichi and {Labbe}, Ivo and {Ma}, Yilun and {Maseda}, Michael V. and {Meyer}, Romain and {Nelson}, Erica and {Oesch}, Pascal and {Xiao}, Mengyuan},
        title = "{The Engine and its Flows: Little Red Dot spectra are shaped by the column densities of their gas envelopes}",
      journal = {arXiv e-prints},
         year = 2026,
        month = mar,
          eid = {arXiv:2603.17667},
        pages = {arXiv:2603.17667},
          doi = {10.48550/arXiv.2603.17667},
archivePrefix = {arXiv},
       eprint = {2603.17667},
 primaryClass = {astro-ph.GA},
       adsurl = {https://ui.adsabs.harvard.edu/abs/2026arXiv260317667M}
}

@ARTICLE{degraaff2025b,
       author = {{de Graaff}, Anna and {Hviding}, Raphael E. and {Naidu}, Rohan P. and {Greene}, Jenny E. and {Miller}, Tim B. and {Leja}, Joel and {Matthee}, Jorryt and {Brammer}, Gabriel and {Katz}, Harley and {Bezanson}, Rachel and {Boogaard}, Leindert A. and {Bose}, Sownak and {Chisholm}, John and {Cleri}, Nikko J. and {Dayal}, Pratika and {Feldmann}, Robert and {Fudamoto}, Yoshinobu and {Fujimoto}, Seiji and {Furtak}, Lukas J. and {Glazebrook}, Karl and {Gottumukkala}, Rashmi and {Heintz}, Kasper E. and {Kokorev}, Vasily and {Labbe}, Ivo and {Maseda}, Michael V. and {McConachie}, Ian and {Nanayakkara}, Themiya and {Nelson}, Erica and {Nowaczyk}, Przemys{\l}aw and {Oesch}, Pascal A. and {Rix}, Hans-Walter and {Setton}, David J. and {Torralba}, Alberto and {Walter}, Fabian and {Wang}, Bingjie and {Weibel}, Andrea and {van der Wel}, Arjen},
        title = "{Little Red Dots host Black Hole Stars: A unified family of gas-reddened AGN revealed by JWST/NIRSpec spectroscopy}",
      journal = {arXiv e-prints},
         year = 2025,
        month = nov,
          eid = {arXiv:2511.21820},
        pages = {arXiv:2511.21820},
          doi = {10.48550/arXiv.2511.21820},
archivePrefix = {arXiv},
       eprint = {2511.21820},
 primaryClass = {astro-ph.GA},
       adsurl = {https://ui.adsabs.harvard.edu/abs/2025arXiv251121820D}
}

@ARTICLE{Nandal2026,
       author = {{Nandal}, Devesh and {Loeb}, Abraham},
        title = "{Supermassive Stars Match the Spectral Signatures of JWST's Little Red Dots}",
      journal = {\apj},
         year = 2026,
        month = feb,
       volume = {998},
       number = {1},
          eid = {124},
        pages = {124},
          doi = {10.3847/1538-4357/ae32f3},
archivePrefix = {arXiv},
       eprint = {2507.12618},
 primaryClass = {astro-ph.GA},
       adsurl = {https://ui.adsabs.harvard.edu/abs/2026ApJ...998..124N}
}

@ARTICLE{Maiolino2025,
       author = {{Maiolino}, Roberto and {Uebler}, Hannah and {D'Eugenio}, Francesco and {Scholtz}, Jan and {Juodzbalis}, Ignas and {Ji}, Xihan and {Perna}, Michele and {Bromm}, Volker and {Dayal}, Pratika and {Koudmani}, Sophie and {Liu}, Boyuan and {Schneider}, Raffaella and {Sijacki}, Debora and {Valiante}, Rosa and {Trinca}, Alessandro and {Zhang}, Saiyang and {Volonteri}, Marta and {Inayoshi}, Kohei and {Carniani}, Stefano and {Nakajima}, Kimihiko and {Isobe}, Yuki and {Witstok}, Joris and {Jones}, Gareth C. and {Tacchella}, Sandro and {Arribas}, Santiago and {Bunker}, Andrew and {Cataldi}, Elisa and {Charlot}, Stephane and {Cresci}, Giovanni and {Curti}, Mirko and {Fabian}, Andrew C. and {Katz}, Harley and {Kumari}, Nimisha and {Laporte}, Nicolas and {Mazzolari}, Giovanni and {Robertson}, Brant and {Sun}, Fengwu and {Rodriguez Del Pino}, Bruno and {Venturi}, Giacomo},
        title = "{A black hole in a near-pristine galaxy 700 million years after the Big Bang}",
      journal = {arXiv e-prints},
         year = 2025,
        month = may,
          eid = {arXiv:2505.22567},
        pages = {arXiv:2505.22567},
          doi = {10.48550/arXiv.2505.22567},
archivePrefix = {arXiv},
       eprint = {2505.22567},
 primaryClass = {astro-ph.GA},
       adsurl = {https://ui.adsabs.harvard.edu/abs/2025arXiv250522567M}
}

@ARTICLE{Zwick2025,
       author = {{Zwick}, Lorenz and {Tiede}, Christopher and {Mayer}, Lucio},
        title = "{Little Red Dots as self-gravitating discs accreting on supermassive stars: Spectral appearance and formation pathway of the progenitors to direct collapse black holes}",
      journal = {arXiv e-prints},
         year = 2025,
        month = jul,
          eid = {arXiv:2507.22014},
        pages = {arXiv:2507.22014},
          doi = {10.48550/arXiv.2507.22014},
archivePrefix = {arXiv},
       eprint = {2507.22014},
 primaryClass = {astro-ph.GA},
       adsurl = {https://ui.adsabs.harvard.edu/abs/2025arXiv250722014Z}
}

@ARTICLE{Chisholm2026,
       author = {{Chisholm}, John and {Berg}, Danielle A. and {Boylan-Kolchin}, Michael and {de Graaff}, Anna and {Furtak}, Lukas J. and {Kokorev}, Vasily and {Matthee}, Jorryt and {Mu{\~n}oz}, Julian B. and {Naidu}, Rohan P. and {Sander}, Andreas A.~C.},
        title = "{Little Red Dots as Globular Clusters in Formation}",
      journal = {arXiv e-prints},
         year = 2026,
        month = feb,
          eid = {arXiv:2602.15935},
        pages = {arXiv:2602.15935},
          doi = {10.48550/arXiv.2602.15935},
archivePrefix = {arXiv},
       eprint = {2602.15935},
 primaryClass = {astro-ph.GA},
       adsurl = {https://ui.adsabs.harvard.edu/abs/2026arXiv260215935C}
}

@ARTICLE{Kiriakidis1993,
       author = {{Kiriakidis}, M. and {Fricke}, K.~J. and {Glatzel}, W.},
        title = "{The stability of massive stars and its dependence on metallicity and opacity.}",
      journal = {\mnras},
         year = 1993,
        month = sep,
       volume = {264},
        pages = {50-62},
          doi = {10.1093/mnras/264.1.50},
       adsurl = {https://ui.adsabs.harvard.edu/abs/1993MNRAS.264...50K}
}

@ARTICLE{Glatzel1994,
       author = {{Glatzel}, W.},
        title = "{On the origin of strange modes and the mechanism of related instabilities}",
      journal = {\mnras},
         year = 1994,
        month = nov,
       volume = {271},
        pages = {66},
          doi = {10.1093/mnras/271.1.66},
       adsurl = {https://ui.adsabs.harvard.edu/abs/1994MNRAS.271...66G}
}

@ARTICLE{Saio1998,
       author = {{Saio}, Hideyuki and {Baker}, Norman H. and {Gautschy}, Alfred},
        title = "{On the properties of strange modes.}",
      journal = {\mnras},
         year = 1998,
        month = mar,
       volume = {294},
        pages = {622-634},
          doi = {10.1111/j.1365-8711.1998.01195.x10.1046/j.1365-8711.1998.01195.x},
       adsurl = {https://ui.adsabs.harvard.edu/abs/1998MNRAS.294..622S}
}

@INPROCEEDINGS{Godart2011,
       author = {{Godart}, M{\'e}lanie and {Dupret}, Marc-Antoine and {Noels}, Arlette and {Aerts}, Conny and {Sim{\'o}n-D{\'\i}az}, Sergio and {Lefever}, Karolien and {Puls}, Joachim and {Montalban}, Josefina and {Ventura}, Paolo},
        title = "{Pulsations in massive stars: effect of the atmosphere on the strange mode pulsations}",
    booktitle = {Active OB Stars: Structure, Evolution, Mass Loss, and Critical Limits},
         year = 2011,
       editor = {{Neiner}, Coralie and {Wade}, Gregg and {Meynet}, Georges and {Peters}, Geraldine},
       series = {IAU Symposium},
       volume = {272},
        month = jul,
        pages = {503-504},
          doi = {10.1017/S1743921311011185},
       adsurl = {https://ui.adsabs.harvard.edu/abs/2011IAUS..272..503G}
}

@INPROCEEDINGS{Saio2013,
       author = {{Saio}, H. and {Georgy}, C. and {Meynet}, G.},
        title = "{Strange-Mode Instability for Micro-Variations in Luminous Blue Variables}",
    booktitle = {Progress in Physics of the Sun and Stars: A New Era in Helio- and Asteroseismology},
         year = 2013,
       editor = {{Shibahashi}, H. and {Lynas-Gray}, A.~E.},
       series = {Astronomical Society of the Pacific Conference Series},
       volume = {479},
        month = dec,
        pages = {47},
          doi = {10.48550/arXiv.1305.4728},
archivePrefix = {arXiv},
       eprint = {1305.4728},
 primaryClass = {astro-ph.SR},
       adsurl = {https://ui.adsabs.harvard.edu/abs/2013ASPC..479...47S}
}

@ARTICLE{Sonoi2014,
       author = {{Sonoi}, Takafumi and {Shibahashi}, Hiromoto},
        title = "{Stability analysis of strange-modes in hot massive stars with time-dependent convection}",
      journal = {\pasj},
         year = 2014,
        month = jul,
       volume = {66},
       number = {4},
          eid = {69},
        pages = {69},
          doi = {10.1093/pasj/psu045},
archivePrefix = {arXiv},
       eprint = {1404.5264},
 primaryClass = {astro-ph.SR},
       adsurl = {https://ui.adsabs.harvard.edu/abs/2014PASJ...66...69S}
}

@ARTICLE{Yadav2018,
       author = {{Yadav}, Abhay Pratap and {K{\"u}hnrich Biavatti}, Stefan Henrique and {Glatzel}, Wolfgang},
        title = "{Strange mode instabilities and mass loss in evolved massive primordial stars}",
      journal = {\mnras},
         year = 2018,
        month = apr,
       volume = {475},
       number = {4},
        pages = {4881-4890},
          doi = {10.1093/mnras/sty092},
       adsurl = {https://ui.adsabs.harvard.edu/abs/2018MNRAS.475.4881Y}
}

@ARTICLE{Inayoshi2013,
       author = {{Inayoshi}, Kohei and {Hosokawa}, Takashi and {Omukai}, Kazuyuki},
        title = "{Pulsational instability of supergiant protostars: do they grow supermassive by accretion?}",
      journal = {\mnras},
         year = 2013,
        month = jun,
       volume = {431},
       number = {4},
        pages = {3036-3044},
          doi = {10.1093/mnras/stt362},
archivePrefix = {arXiv},
       eprint = {1302.6065},
 primaryClass = {astro-ph.SR},
       adsurl = {https://ui.adsabs.harvard.edu/abs/2013MNRAS.431.3036I}
}

@ARTICLE{Nakauchi2020,
       author = {{Nakauchi}, Daisuke and {Inayoshi}, Kohei and {Omukai}, Kazuyuki},
        title = "{Pulsation-driven Mass Loss from Massive Stars behind Stellar Mergers in Metal-poor Dense Clusters}",
      journal = {\apj},
         year = 2020,
        month = oct,
       volume = {902},
       number = {1},
          eid = {81},
        pages = {81},
          doi = {10.3847/1538-4357/abb463},
archivePrefix = {arXiv},
       eprint = {2008.13647},
 primaryClass = {astro-ph.SR},
       adsurl = {https://ui.adsabs.harvard.edu/abs/2020ApJ...902...81N}
}

@ARTICLE{Saio2024,
       author = {{Saio}, Hideyuki and {Nandal}, Devesh and {Ekstr{\"o}m}, Sylvia and {Meynet}, George},
        title = "{Linear adiabatic analysis for general-relativistic instability in primordial accreting supermassive stars}",
      journal = {\aap},
         year = 2024,
        month = sep,
       volume = {689},
          eid = {A169},
        pages = {A169},
          doi = {10.1051/0004-6361/202449971},
archivePrefix = {arXiv},
       eprint = {2406.18040},
 primaryClass = {astro-ph.HE},
       adsurl = {https://ui.adsabs.harvard.edu/abs/2024A&A...689A.169S}
}

@ARTICLE{Nandal2025,
       author = {{Nandal}, Devesh and {Whalen}, Daniel J. and {Latif}, Muhammad A. and {Heger}, Alexander},
        title = "{1000─10,000 M$_{{\ensuremath{\odot}}}$ Primordial Stars Created the Nitrogen Excess in GS 3073 at z = 5.55}",
      journal = {\apjl},
         year = 2025,
        month = nov,
       volume = {994},
       number = {1},
          eid = {L11},
        pages = {L11},
          doi = {10.3847/2041-8213/ae1a63},
archivePrefix = {arXiv},
       eprint = {2502.04435},
 primaryClass = {astro-ph.GA},
       adsurl = {https://ui.adsabs.harvard.edu/abs/2025ApJ...994L..11N}
}

@ARTICLE{Nandal2025b,
       author = {{Nandal}, Devesh and {Buldgen}, Ga{\"e}l and {Whalen}, Daniel J. and {Regan}, John and {Woods}, Tyrone E. and {Tan}, Jonathan C.},
        title = "{Rotating supermassive Pop III stars on the main sequence}",
      journal = {\aap},
         year = 2025,
        month = sep,
       volume = {701},
          eid = {A262},
        pages = {A262},
          doi = {10.1051/0004-6361/202555878},
archivePrefix = {arXiv},
       eprint = {2506.08268},
 primaryClass = {astro-ph.SR},
       adsurl = {https://ui.adsabs.harvard.edu/abs/2025A&A...701A.262N}
}

@ARTICLE{Haemmerle2021,
       author = {{Haemmerl{\'e}}, L.},
        title = "{General-relativistic instability in rapidly accreting supermassive stars: The impact of rotation}",
      journal = {\aap},
         year = 2021,
        month = jun,
       volume = {650},
          eid = {A204},
        pages = {A204},
          doi = {10.1051/0004-6361/202140893},
archivePrefix = {arXiv},
       eprint = {2104.11754},
 primaryClass = {astro-ph.HE},
       adsurl = {https://ui.adsabs.harvard.edu/abs/2021A&A...650A.204H}
}

@BOOK{Levesque17,
   author = {{Levesque}, Emily M.},
        title = {Astrophysics of Red Supergiants},
    publisher = {Institute of Physics},
         year = 2017,
          doi = {10.1088/978-0-7503-1329-2},
       adsurl = {https://ui.adsabs.harvard.edu/abs/2017ars..book.....L}
}

@ARTICLE{2025ApJ...987..203G,
   author = {{Guhathakurta}, Puragra and {Grion Filho}, Douglas and {Bhattacharya}, Antara R. and {Cullinane}, Lara R. and {Dalcanton}, Julianne J. and {Gilbert}, Karoline M. and {Girardi}, Leo and {Kamath}, Anika and {Kirby}, Evan N. and {Maheshwari}, Arya and {Marigo}, Paola and {Masegian}, Alexand ra and {Quirk}, Amand a C.~N. and {Raikar}, Rachel and {Rinehart}, Stanley M. and {Rodriguez}, Caelum J. and {Williams}, Benjamin F.},
        title = {Discovery of a Weak CN Spectral Absorption Feature in Red Supergiant Stars in the Andromeda (M31) and Triangulum (M33) Galaxies},
      journal = {\apj},
         year = 2025,
        month = jul,
       volume = {987},
       number = {2},
          eid = {203},
        pages = {203},
          doi = {10.3847/1538-4357/adcc2c},
archivePrefix = {arXiv},
       eprint = {2504.04291},
 primaryClass = {astro-ph.SR},
       adsurl = {https://ui.adsabs.harvard.edu/abs/2025ApJ...987..203G}
}

@ARTICLE{2026MNRAS.545f2235J,
   author = {{Ji}, Xihan and {D'Eugenio}, Francesco and {Juod{\v{z}}balis}, Ignas and {Walton}, Dominic J. and {Fabian}, Andrew C. and {Maiolino}, Roberto and {Ramos Almeida}, Cristina and {Acosta Pulido}, Jose A. and {Belokurov}, Vasily A. and {Isobe}, Yuki and {Jones}, Gareth and {Maraston}, Claudia and {Scholtz}, Jan and {Simmonds}, Charlotte and {Tacchella}, Sand ro and {Terlevich}, Elena and {Terlevich}, Roberto},
        title = {Lord of LRDs: insights into a 'Little Red Dot' with a low-ionization spectrum at z = 0.1},
      journal = {\mnras},
         year = 2026,
        month = jan,
       volume = {545},
       number = {3},
          eid = {staf2235},
        pages = {staf2235},
          doi = {10.1093/mnras/staf2235},
archivePrefix = {arXiv},
       eprint = {2507.23774},
 primaryClass = {astro-ph.GA},
       adsurl = {https://ui.adsabs.harvard.edu/abs/2026MNRAS.545f2235J}
}

@ARTICLE{2025ApJ...980L..34L,
   author = {{Lin}, Ruqiu and {Zheng}, Zhen-Ya and {Jiang}, Chunyan and {Yuan}, Fang-Ting and {Ho}, Luis C. and {Wang}, Junxian and {Jiang}, Linhua and {Rhoads}, James E. and {Malhotra}, Sangeeta and {Barrientos}, L. Felipe and {Wold}, Isak and {Infante}, Leopoldo and {Zhu}, Shuairu and {Ji}, Xiang and {Fu}, Xiaodan},
        title = {Discovery of Local Analogs to JWST's Little Red Dots},
      journal = {\apjl},
         year = 2025,
        month = feb,
       volume = {980},
       number = {2},
          eid = {L34},
        pages = {L34},
          doi = {10.3847/2041-8213/adaaf1},
archivePrefix = {arXiv},
       eprint = {2412.08396},
 primaryClass = {astro-ph.GA},
       adsurl = {https://ui.adsabs.harvard.edu/abs/2025ApJ...980L..34L}
}

@ARTICLE{Liu2025,
       author = {{Liu}, Hanpu and {Jiang}, Yan-Fei and {Quataert}, Eliot and {Greene}, Jenny E. and {Ma}, Yilun},
        title = "{The Balmer Break and Optical Continuum of Little Red Dots from Super-Eddington Accretion}",
      journal = {\apj},
         year = 2025,
        month = nov,
       volume = {994},
       number = {1},
          eid = {113},
        pages = {113},
          doi = {10.3847/1538-4357/ae0c19},
archivePrefix = {arXiv},
       eprint = {2507.07190},
 primaryClass = {astro-ph.GA},
       adsurl = {https://ui.adsabs.harvard.edu/abs/2025ApJ...994..113L}
}

@ARTICLE{Kido2025,
       author = {{Kido}, Daisaburo and {Ioka}, Kunihito and {Hotokezaka}, Kenta and {Inayoshi}, Kohei and {Irwin}, Christopher M.},
        title = "{Black hole envelopes in Little Red Dots}",
      journal = {\mnras},
         year = 2025,
        month = dec,
       volume = {544},
       number = {4},
        pages = {3407-3416},
          doi = {10.1093/mnras/staf1898},
archivePrefix = {arXiv},
       eprint = {2505.06965},
 primaryClass = {astro-ph.HE},
       adsurl = {https://ui.adsabs.harvard.edu/abs/2025MNRAS.544.3407K}
}

@ARTICLE{Wang2026,
       author = {{Wang}, Bingjie and {Leja}, Joel and {Labbe}, Ivo and {Greene}, Jenny E. and {Liu}, Hanpu and {de Graaff}, Anna and {Hviding}, Raphael E. and {Matthee}, Jorryt and {Quataert}, Eliot and {Bezanson}, Rachel and {Boogaard}, Leindert A. and {Brammer}, Gabriel and {Burgasser}, Adam J. and {Chen}, Yi-Xian and {Cleri}, Nikko J. and {Cutler}, Sam E. and {Dayal}, Pratika and {Furtak}, Lukas J. and {Fujimoto}, Seiji and {Glazebrook}, Karl and {Goulding}, Andy D. and {Helton}, Jakob M. and {Hirschmann}, Michaela and {Jiang}, Yan-Fei and {Kokorev}, Vasily and {Ma}, Yilun and {Miller}, Tim B. and {Naidu}, Rohan P. and {Oesch}, Pascal and {Pan}, Richard and {Papovich}, Casey and {Price}, Sedona H. and {Rix}, Hans-Walter and {Setton}, David J. and {Sun}, Wendy Q. and {Weaver}, John R. and {Whitaker}, Katherine E. and {Zitrin}, Adi},
        title = "{Water absorption confirms cool atmospheres in two little red dots}",
      journal = {arXiv e-prints},
         year = 2026,
        month = feb,
          eid = {arXiv:2602.06024},
        pages = {arXiv:2602.06024},
          doi = {10.48550/arXiv.2602.06024},
archivePrefix = {arXiv},
       eprint = {2602.06024},
 primaryClass = {astro-ph.GA},
       adsurl = {https://ui.adsabs.harvard.edu/abs/2026arXiv260206024W}
}

@ARTICLE{Humphreys2002,
       author = {{Humphreys}, Roberta M. and {Davidson}, Kris and {Smith}, Nathan},
        title = "{Crossing the Yellow Void: Spatially Resolved Spectroscopy of the Post-Red Supergiant IRC +10420 and Its Circumstellar Ejecta}",
      journal = {\aj},
         year = 2002,
        month = aug,
       volume = {124},
       number = {2},
        pages = {1026-1044},
          doi = {10.1086/341380},
archivePrefix = {arXiv},
       eprint = {astro-ph/0205247},
 primaryClass = {astro-ph},
       adsurl = {https://ui.adsabs.harvard.edu/abs/2002AJ....124.1026H}
}

@ARTICLE{Humphreys1997,
       author = {{Humphreys}, Roberta M. and {Smith}, Nathan and {Davidson}, Kris and {Jones}, Terry Jay and {Gehrz}, Robert T. and {Mason}, Christopher G. and {Hayward}, Thomas L. and {Houck}, James R. and {Krautter}, Joachim},
        title = "{HST and Infrared Images of the Circumstellar Environment of the Cool Hypergiant IRC + 10420}",
      journal = {\aj},
         year = 1997,
        month = dec,
       volume = {114},
        pages = {2778},
          doi = {10.1086/118686},
       adsurl = {https://ui.adsabs.harvard.edu/abs/1997AJ....114.2778H}
}

@ARTICLE{Lagadec2011,
       author = {{Lagadec}, E. and {Zijlstra}, A.~A. and {Oudmaijer}, R.~D. and {Verhoelst}, T. and {Cox}, N.~L.~J. and {Szczerba}, R. and {M{\'e}karnia}, D. and {van Winckel}, H.},
        title = "{A double detached shell around a post-red supergiant: IRAS 17163-3907, the Fried Egg nebula}",
      journal = {\aap},
         year = 2011,
        month = oct,
       volume = {534},
          eid = {L10},
        pages = {L10},
          doi = {10.1051/0004-6361/201117521},
archivePrefix = {arXiv},
       eprint = {1109.5947},
 primaryClass = {astro-ph.SR},
       adsurl = {https://ui.adsabs.harvard.edu/abs/2011A&A...534L..10L}
}

@ARTICLE{Wallstrom2015,
       author = {{Wallstr{\"o}m}, S.~H.~J. and {Muller}, S. and {Lagadec}, E. and {Black}, J.~H. and {Oudmaijer}, R.~D. and {Justtanont}, K. and {van Winckel}, H. and {Zijlstra}, A.~A.},
        title = "{Investigating the nature of the Fried Egg nebula. CO mm-line and optical spectroscopy of IRAS 17163-3907}",
      journal = {\aap},
         year = 2015,
        month = feb,
       volume = {574},
          eid = {A139},
        pages = {A139},
          doi = {10.1051/0004-6361/201321516},
archivePrefix = {arXiv},
       eprint = {1501.03362},
 primaryClass = {astro-ph.SR},
       adsurl = {https://ui.adsabs.harvard.edu/abs/2015A&A...574A.139W}
}

\appendix
\setcounter{table}{0}
\renewcommand{\thetable}{A\arabic{table}}

\section{Extended Methods}
\label{app:extended_methods}

We analyse the full evolutionary outputs of five \textsc{GENEC} supermassive-star sequences. Each stored model is treated as a complete stellar structure at one evolutionary time: we read the full radial profile from the \texttt{StrucData} file and use the auxiliary \texttt{v} file, when present, to supply additional thermodynamic and transport quantities. The calculation is therefore performed on the full star for every stored model, rather than on a reduced envelope subset.

For each model we retain the radial profiles needed for the pulsation analysis,
\[
r,\quad M_r,\quad P,\quad \rho,\quad T,\quad \kappa,\quad \epsilon,\quad
L_{\rm rad},\quad L_{\rm tot},
\]
together with composition, convective quantities, and the thermodynamic derivatives used to evaluate the adiabatic exponents,
\begin{align}
\Gamma_1 &= \frac{\chi_\rho}{1-\chi_T \nabla_{\rm ad}},\\
\Gamma_3-1 &= \Gamma_1 \nabla_{\rm ad}.
\end{align}
Before proceeding, we require the structure to remain physically well behaved, including monotonic radius and enclosed mass and finite thermodynamic variables.

\subsection{Candidate envelope regions}

The later nonadiabatic analysis is not carried out blindly over the whole star. Instead, we first identify physically motivated envelope regions where radial driving is most likely to occur. These include the H, He\,I, and He\,II partial-ionization zones, the iron-opacity bump, the deeper iron feature, and a near-surface radiative layer \citep{Saio2024}. In practice we use the temperature intervals
\begin{align}
{\rm H:}\quad &3.75 \le \log T \le 4.20,\\
{\rm He\,I:}\quad &4.20 \le \log T \le 4.75,\\
{\rm He\,II:}\quad &4.55 \le \log T \le 5.10,\\
{\rm Fe:}\quad &5.15 \le \log T \le 5.35,\\
{\rm deep\ Fe:}\quad &6.15 \le \log T \le 6.40,
\end{align}
together with an outer surface layer defined by \(q=M_r/M_\star \gtrsim 0.99\).

We also evaluate simple envelope diagnostics associated with strange-mode behaviour in luminous, radiation-dominated stars, including large \(L/M\), low outer gas-pressure fraction, short local thermal times, and a large radiative flux fraction \citep{Saio1998,Saio2013,Sonoi2014}. These quantities are used only to flag plausible driving regions. They do not determine the final mode selection, which is based instead on the shell-by-shell work analysis described below.

\subsection{Radial modes}

We begin with linear adiabatic radial pulsations. In Newtonian gravity, writing the radial displacement as \(\xi_r=r\eta\), the pulsation equation may be written in self-adjoint form,
\begin{equation}
\frac{d}{dr}\!\left[A(r)\frac{d\eta}{dr}\right]
+
\left[B(r)+\omega^2 C(r)\right]\eta
=0,
\label{eq:app_SL}
\end{equation}
with
\begin{align}
A(r) &= \Gamma_1 P r^4,\\
B(r) &= r^3 \frac{d}{dr}\!\left[(3\Gamma_1-4)P\right],\\
C(r) &= \rho r^4.
\end{align}
Regularity is imposed at the centre, and the outer boundary condition is the vanishing of the Lagrangian pressure perturbation,
\begin{equation}
\Delta P = 0.
\end{equation}
We solve for the lowest radial modes by scanning eigenfrequency space, locating sign changes in the surface residual, and refining each root by bisection.

To place the same sequences in a relativistic context, we also solve the general-relativistic radial problem following \citet{Saio2024}. The equilibrium metric is written as
\begin{equation}
ds^2 = -e^{2a} c^2 dt^2 + e^{2b} dr^2 + r^2 d\Omega^2,
\end{equation}
and the perturbation equations are solved in the variables
\begin{equation}
Y_1=\frac{\xi_r}{r},
\qquad
Y_2=\frac{\Delta P}{P},
\end{equation}
with
\begin{align}
\frac{dY_1}{d\ln r} &=
-\left(3-\frac{da}{d\ln r}\right)Y_1
-\frac{Y_2}{\Gamma_1},\\
\frac{dY_2}{d\ln r} &=
\left[A_{\rm GR}+\omega^2 D_{\rm GR}\right]Y_1
+B_{\rm GR}Y_2.
\end{align}
This GR calculation is used to follow the approach to relativistic softening and instability. The Newtonian modes remain the default basis for the driving and mass-loss estimates described below.

\subsection{Quasi-nonadiabatic driving and mode selection}

The next step is not a full nonadiabatic eigenvalue calculation. Instead, we evaluate local driving and damping terms on the adiabatic eigenfunctions. This provides a transparent picture of where positive work is concentrated and which physical channel is responsible for it.

The local compression implied by the radial displacement is
\begin{equation}
\Delta \ln \rho
\approx
-\frac{1}{r^2}\frac{d}{dr}\!\left(r^2 \xi_r\right),
\label{eq:app_dlnrho}
\end{equation}
and the associated adiabatic temperature response is
\begin{equation}
\Delta \ln T_{\rm ad}
=
(\Gamma_3-1)\,\Delta \ln \rho.
\label{eq:app_dlnT}
\end{equation}

To measure how effectively a layer can exchange heat over one cycle, we compare the pulsation period \(P\) with the local thermal time,
\begin{equation}
t_{\rm th}(r_i)
\approx
\frac{\sum_{j=i}^{N} C_{V,j} T_j \Delta m_j}{L_\star},
\end{equation}
and define
\begin{equation}
x_{\rm th}=\frac{P}{t_{\rm th}},
\qquad
f_{\rm th}=\frac{2x_{\rm th}}{1+x_{\rm th}^2}.
\end{equation}
This weighting peaks when \(P\sim t_{\rm th}\). Convective damping is treated in the same spirit by comparing the pulsation period with the local convective turnover time.

Opacity driving is measured through the response of the radiative conductivity,
\begin{equation}
K_{\rm rad}\propto \frac{T^3}{\kappa \rho},
\end{equation}
to adiabatic compression. Using
\begin{equation}
d\ln\kappa
=
\kappa_T\, d\ln T + \kappa_\rho\, d\ln\rho,
\end{equation}
together with the adiabatic relation for \(d\ln T\), we obtain
\begin{equation}
\left(\frac{d\ln K_{\rm rad}}{d\ln\rho}\right)_{\rm ad}
=
(3-\kappa_T)(\Gamma_3-1) - (1+\kappa_\rho).
\label{eq:app_trapcoeff}
\end{equation}
Compression increases radiative trapping when this quantity is negative, so the sign-reversed form serves as our local opacity-driving proxy.

For completeness we also monitor the adiabatic sensitivity of the nuclear energy generation rate,
\begin{equation}
\epsilon_{\rm coeff}
=
\epsilon_T + (\Gamma_3-1)\epsilon_\rho.
\label{eq:app_epscoeff}
\end{equation}
This term is retained as a diagnostic, but it is not included in the default net-driving sum and is not used as the primary criterion for mode selection.

To represent the outer-envelope contribution, we include a separate surface term that grows when the local layer is radiatively dominated, weakly supported by gas pressure, close to the surface, and dynamically important in the eigenfunction. We interpret this term as a proxy for surface-leakage or strange-mode-like behaviour rather than as a formal strange-mode solution.

The local power densities are therefore written schematically as
\begin{align}
\dot p_\kappa &\propto
\frac{\rho C_V T}{t_{\rm th}}
\left(\Delta\ln T_{\rm ad}\right)^2
f_{\rm rad}\,f_{\rm th}\,\mathcal{T}_\kappa,\\
\dot p_{\rm surf} &\propto
\frac{\rho C_V T}{t_{\rm th}}
\left(\Delta\ln T_{\rm ad}\right)^2
\mathcal{S}_{\rm surf},\\
\dot p_{\rm conv} &\propto
-
\frac{\rho C_V T}{t_{\rm th}}
\left(\Delta\ln T_{\rm ad}\right)^2
f_{\rm conv},
\end{align}
where \(f_{\rm rad}=L_{\rm rad}/L_{\rm tot}\) after repair, \(\mathcal{T}_\kappa\) is the opacity-trapping factor, and \(\mathcal{S}_{\rm surf}\) is the surface-driving proxy. The default net driving is
\begin{equation}
\dot p_{\rm tot}
=
\dot p_\kappa + \dot p_{\rm surf} + \dot p_{\rm conv}.
\label{eq:app_pdot_total}
\end{equation}

From these local terms we form the global mode energy and driving power,
\begin{align}
E_{\rm mode} &=
\int \frac{1}{2}\rho \omega_r^2 \xi_r^2\, dV,\\
\dot P &=
\int \dot p_{\rm tot}\, dV,
\end{align}
and define the corresponding growth proxy,
\begin{equation}
\gamma=\frac{\dot P}{2E_{\rm mode}}.
\label{eq:app_gamma}
\end{equation}
The work contributed by each shell over one cycle is
\begin{equation}
W_{{\rm shell},i}
=
\dot p_i\,\Delta V_i\,P.
\end{equation}

The mode carried forward at each stored model is chosen by balancing three considerations: positive growth, concentration of positive work in a physically plausible envelope layer, and consistency between that work-producing layer and the candidate region identified from the equilibrium structure.

\subsection{Re-identifying the driving layer}

The outermost zones of luminous stellar models can contain numerically unreliable radiative fractions \citep{Maeder2000}. For that reason, we re-examine the candidate driving region before converting any unstable mode into a mass-loss estimate.

We first rewrite the opacity derivatives in more diagnostic forms. At constant pressure,
\begin{equation}
\left(\frac{\partial \ln \kappa}{\partial \ln T}\right)_P
=
\kappa_T - \frac{\chi_T}{\chi_\rho}\kappa_\rho,
\label{eq:app_kconstP}
\end{equation}
while along an adiabatic path,
\begin{equation}
\left(\frac{d\ln\kappa}{d\ln T}\right)_{\rm ad}
=
\kappa_T + \frac{\kappa_\rho}{\Gamma_3-1}.
\label{eq:app_kad}
\end{equation}
These combinations provide compact measures of whether compression tends to increase radiative trapping.

Where the raw radiative luminosity fraction becomes unphysical, we reconstruct the radiative luminosity from the diffusion form of the temperature gradient,
\begin{equation}
L_{\rm rad,rec}
=
\frac{16\pi a c G M_r T^4 \nabla}{3\kappa P}.
\label{eq:app_lradrec}
\end{equation}
We use the original profile where it remains admissible and substitute the reconstructed value only in problematic cells. The repaired profile is then used to rank the candidate driving zones.

This step does not redefine the stellar model. Its purpose is simply to ensure that the inferred driving layer is physically plausible and not an artifact of a few pathological outer mesh points.

\subsection{Caveat on steady radiative mass loss}

Other forms of mass loss may also operate during SMS evolution and are not excluded here. During brief hot blueward excursions, line-driven winds of the type described by \citet{Vink2001} could in principle contribute, while on the cool supergiant side one may appeal to empirical prescriptions such as \citet{deJager1988}. The difficulty is that these prescriptions are not calibrated for near-Eddington SMS envelopes.

This uncertainty was explored by \citet{Nandal2026b} for a $Z=10^{-4}\,Z_\odot$ SMS. Starting from a mid core-H-burning model of $7.3973\times10^4\,M_\odot$, they evolved the star without further accretion or collisions until H exhaustion. The de Jager prescription reduced the final mass to $2.3421\times10^4\,M_\odot$, implying a total loss of $\sim5.1\times10^4\,M_\odot$, whereas the Vink prescription gave $M_f=6.2348\times10^4\,M_\odot$, implying a loss of only $\sim1.2\times10^4\,M_\odot$. In that test, the Vink-like wind became important only during brief hot contraction phases with $T_{\rm eff}\gtrsim2.5\times10^4\,{\rm K}$. These values should therefore be read only as first-order bounds rather than as robust SMS wind predictions. For that reason, and to isolate the role of discrete pulsational ejection, we do not impose an additional steady radiative-wind prescription in the fiducial calculations. The mass loss discussed here should therefore be read specifically as the pulsation-linked component.

\subsection{From unstable modes to mass loss}

Once a driving layer has been identified, we translate the pulsation diagnostics into a conservative estimate of pulsation-linked mass loss. The coupling is written in terms of a luminosity fraction,
\begin{equation}
\begin{split}
f_{\rm coup} =
\min\!\Bigl[
&f_{{\rm coup},\max}, \\
& C_{\rm branch}\, g_P\,
\max\!\left(W_{\rm metric},\,O_{\rm metric}\right)\,
w_{\rm conf}\, b_{\rm ch}
\Bigr]
\end{split}
\label{eq:app_fcoup}
\end{equation}

where \(g_P=\gamma P\) is the growth per period, \(W_{\rm metric}\) and \(O_{\rm metric}\) measure how strongly the inferred driving layer participates in the work budget, \(w_{\rm conf}\) is a confidence weight, and \(b_{\rm ch}\) allows a modest dependence on the dominant driving channel. The branch-dependent ceiling \(f_{{\rm coup},\max}\) is listed in Table~\ref{tab:A1_branch_params}. We define
\begin{equation}
\begin{split}
W_{\rm metric} &= (f_+ f_{|W|})^{1/2}, \\
O_{\rm metric} &= 0.6 f_+ + 0.4 \min(1,\,2f_{|W|}) .
\end{split}
\end{equation}
where \(f_+\) is the fraction of the positive work budget arising in the inferred driving layer and \(f_{|W|}\) is the corresponding absolute-work fraction.

To reflect the weak binding of the outer envelope, we replace the structural escape speed by an effective escape scale,
\begin{equation}
v_{\rm esc,eff}
=
v_{\rm esc}\sqrt{f_{\rm bind,eff}},
\qquad
v_{\rm esc}
=
\left(\frac{2GM_\star}{R_\star}\right)^{1/2},
\label{eq:app_vesceff}
\end{equation}
where \(f_{\rm bind,eff}\) is a reduced binding factor derived from the local gas-pressure fraction and radiative support. This is a simple envelope-binding proxy rather than a full dynamical calculation.

The mass-loss rate is then written in energy-limited form,
\begin{equation}
\dot M
=
\frac{2\eta L_{\rm drive}}{v_{\rm esc,eff}^2},
\qquad
L_{\rm drive}=f_{\rm coup}L_\star.
\label{eq:app_mdot}
\end{equation}
Here \(\eta\) is an explicit efficiency factor, varied over low, fiducial, and high branches. Each estimate is capped by the mass accessible above the inferred driving layer, so that the procedure cannot eject more material than is locally available.

The numerical coefficients used in this mapping are collected in Table~\ref{tab:A1_branch_params}. Surface-dominated cases are allowed to couple more strongly than deeper opacity-driven cases, reflecting the weaker binding of the outermost layers.

\begin{deluxetable*}{lcccc}
\tabletypesize{\scriptsize}
\tablecaption{Branch-dependent parameters used in the pulsation--mass-loss mapping.}
\label{tab:A1_branch_params}
\tablehead{
\colhead{Branch family} &
\colhead{\(C_{\rm branch}\)} &
\colhead{\(f_{{\rm coup},\max}\)} &
\colhead{\(\eta_{\rm low},\,\eta_{\rm fid},\,\eta_{\rm high}\)} &
\colhead{Accessible mass cap \(M_{\rm acc}\)}
}
\startdata
Surface-leakage dominated &
6 &
0.05 &
\(0.10,\ 0.30,\ 1.00\) &
\(\min\!\left[M_{>},\,\max(0.5M_{>},M_{\rm sh})\right]\) \\
Opacity-dominated &
4 &
0.02 &
\(0.03,\ 0.10,\ 0.30\) &
\(\min\!\left(0.35M_{\rm sh},\,M_{>}\right)\) \\
Mixed &
5 &
0.03 &
\(0.05,\ 0.15,\ 0.50\) &
\(\min\!\left(0.6M_{>},\,0.6M_{\rm sh}+0.4M_{>}\right)\) \\
Indeterminate &
3 &
0.015 &
\(0.03,\ 0.10,\ 0.30\) &
\(\min\!\left(0.25M_{\rm sh},\,M_{>}\right)\) \\
\enddata
\tablecomments{\(M_{\rm sh}\) is the mass contained in the inferred driving shell and \(M_{>}\) is the mass above the inner edge of that shell.}
\end{deluxetable*}

This part of the method is necessarily approximate. The present work does not solve nonlinear radiation hydrodynamics, so the mapping from pulsational driving to \(\dot M\) should be read as a structured estimate rather than as a unique prediction.

\subsection{Finite ejection episodes along each sequence}

A non-zero \(\dot M\) at a single stored model is not interpreted as a long-lived steady wind. Instead, it marks that stored model as pulsationally active. We do not impose an additional floor in event mass or duration at this grouping stage: models with \(\dot M=0\) are treated as quiescent, while contiguous active models are merged into one episode. If only one stored model in a local interval is active, we still assign it a finite duration based on the shorter of the local evolutionary spacing and an intrinsic pulsation timescale,
\begin{equation}
\Delta t_{\rm eff}
=
\min\!\left[\Delta t_{\rm half-gap},\,
\max\!\left(N_{\rm grow}\tau_{\rm grow},\,N_P P\right)\right].
\end{equation}
The mass associated with an isolated episode is then
\begin{equation}
\Delta M_{\rm ep}
\simeq
\dot M\,\Delta t_{\rm eff},
\end{equation}
while multi-model episodes are integrated over the sampled active interval. In this bookkeeping, every non-zero \(\dot M\) estimate in the sampled sequence belongs to exactly one episode, so there is no separate off-episode contribution omitted from the cumulative ejecta mass. In all cases, the low, fiducial, high, and upper-limit branches are kept mutually consistent and are subject to the same reservoir cap.

\subsection{Velocity scales and shell estimates}

Finally, we translate the episode-integrated results into a small set of observer-facing quantities. Alongside the escape speed, we define two characteristic pulsation speed scales,
\begin{equation}
v_{R/P}=\frac{R_\star}{P},
\qquad
v_{2\pi R/P}=\frac{2\pi R_\star}{P},
\end{equation}
and the displacement amplitude required for a sinusoidal oscillation to reach escape,
\begin{equation}
A_{\rm crit,esc}
=
\frac{v_{\rm esc}P}{2\pi R_\star}.
\end{equation}
We also report radiative momentum and energy limits,
\begin{equation}
v_{L/c}=\frac{L_\star}{\dot M c},
\qquad
v_{\rm tir}
=
\left[
\max\!\left(0,\frac{2L_\star}{\dot M}-v_{\rm esc}^2\right)
\right]^{1/2}.
\end{equation}
These are not predictions of a unique terminal velocity; they are compact measures of whether the inferred event-averaged mass flux is broadly compatible with the available radiative budget.

To estimate where ejected material would lie at a later time, we propagate each episode outward with a simple bracket of launch speeds,
\begin{equation}
v_{\rm low}=v_{R/P},
\qquad
v_{\rm high}=v_{2\pi R/P}.
\end{equation}
For an episode evaluated at time \(t_{\rm ref}\), with start and end times \(t_{\rm start}\) and \(t_{\rm end}\), the inner and outer shell radii are
\begin{align}
r_{\rm in}
&=
\max\!\left[
r_{\rm launch},
v_{\rm low}(t_{\rm ref}-t_{\rm end})_+
\right],\\
r_{\rm out}
&=
\max\!\left[
r_{\rm in},
r_{\rm launch}+v_{\rm high}(t_{\rm ref}-t_{\rm start})_+
\right],
\end{align}
where \((x)_+=\max(x,0)\). From these we define the shell thickness,
\begin{equation}
\Delta R = r_{\rm out}-r_{\rm in},
\end{equation}
and the thin-shell surface density,
\begin{equation}
\Sigma_{\rm sh}(r)
\simeq
\frac{M_{\rm sh}}{4\pi r^2},
\end{equation}
which gives an optical-depth proxy,
\begin{equation}
\tau(r) \simeq \kappa\,\Sigma_{\rm sh}(r).
\end{equation}
For the shell diagnostics quoted in the main text, we evaluate this proxy at the inner edge, outer edge, and geometric-mean radius of the shell, giving \(\tau_{\rm in}=\tau(r_{\rm in})\), \(\tau_{\rm out}=\tau(r_{\rm out})\), and \(\tau_{\rm geo}=\tau(\sqrt{r_{\rm in}r_{\rm out}})\). Unless stated otherwise, we adopt a constant electron-scattering opacity \(\kappa=0.34\,{\rm cm^2\,g^{-1}}\) for these order-of-magnitude estimates.
\subsection{Composition of the ejected layers}

For selected episodes we also examine the composition of the outer layers implicated by the inferred mass loss. If an episode ejects a mass \(\Delta M_{\rm ep}\), the corresponding outer mass depth is
\begin{equation}
\Delta q_{\rm ep} = \frac{\Delta M_{\rm ep}}{M_\star}.
\end{equation}
We then inspect the abundance profiles over the outer mass interval associated with that depth. This procedure does not introduce any additional mixing or nucleosynthesis; it simply identifies the composition of the layers that would be sampled by an ejection of the inferred depth. These abundance diagnostics are used only as a post-processing aid when interpreting the composition of the expelled material. The abundance ratios reported in the main text are computed from the integrated shell masses over that selected interval and are quoted as shell-averaged logarithmic number ratios.

\subsection{Metallicity dependence of the shell and ejecta diagnostics}

Table~\ref{tab:A2} shows that all five Phase XVI sequences end with nearly the same final mass, \(M_{\rm f}\simeq10^5\,M_\odot\), but not with the same late-time pulsational behaviour. The main metallicity dependence appears instead in the number of eruptive episodes, the amount of mass removed, the compactness of the terminal envelope, and the chemistry of the expelled shell. The central result is already clear from the table itself: pulsation-driven shell ejection is not unique to the \(Z=10^{-2}\,Z_\odot\) model, but is present from Pop III to \(10^{-2}\,Z_\odot\).

The non-monotonic behaviour with metallicity is physically expected. In luminous stars, strange-mode instability is not controlled by a single opacity feature. Classical stability calculations identify one family associated with He ionization and another associated with heavy-element opacity enhancement \citep{Kiriakidis1993}. The first can remain important even at very low metallicity, while the second becomes stronger as metal opacity increases. Near-Eddington envelopes can therefore remain susceptible to inflation and strange-mode driving across a wide metallicity range. The existence of pulsations is thus generic, but their timing, multiplicity, and ejecta yield need not vary smoothly with \(Z\).

This interpretation helps explain the burst statistics in Table~\ref{tab:A2}. The \(Z=10^{-3}\,Z_\odot\) sequence produces the largest number of fiducial episodes, \(N_{\rm ep}=28\), whereas the \(Z=10^{-5}\,Z_\odot\) model yields the largest cumulative ejecta mass, \(\Delta M_{\rm ej}=6.26\times10^3\,M_\odot\) on the fiducial branch and \(1.47\times10^4\,M_\odot\) on the upper branch. By contrast, the \(Z=10^{-2}\,Z_\odot\) case shows only four fiducial episodes and the smallest total ejected mass, \(\Delta M_{\rm ej}=4.80\times10^2\,M_\odot\). Frequent bursting and efficient mass removal are therefore not the same thing.

The terminal structural quantities in Table~\ref{tab:A2} divide the grid into two broad regimes. The Pop III model is the clear outlier: it ends much hotter and more compact, with \(\log T_{{\rm eff},f}=4.872\) and \(v_{{\rm esc,last}}=7731\,{\rm km\,s^{-1}}\). The metal-enriched models instead cluster near \(\log T_{{\rm eff},f}\simeq4.0\) and \(v_{{\rm esc,last}}\simeq1200\)–\(1600\,{\rm km\,s^{-1}}\), while all retain \(\Gamma_{{\rm Edd},f}\approx1\). This suggests that the decisive metallicity effect is not a large change in \(L/M\) itself, but a change in how that near-Eddington luminosity is expressed in the envelope structure. Once the metal-enriched models settle into cooler and less tightly bound envelopes, strange-mode driving can more easily expel optically important outer material.

The shell quantities strengthen the same conclusion. Across the entire grid, the shell identified in Table~\ref{tab:A2} remains compact and optically thick, with \(\tau_{\rm es}\) between \(\sim7.4\times10^2\) and \(1.3\times10^4\). Thus, the production of a dense inner shell is not restricted to one metallicity. What changes with \(Z\) is the balance between total ejecta mass, shell compactness, and terminal stellar state. In this sense, the \(Z=10^{-2}\,Z_\odot\) model is not special because it maximizes the ejected mass. It is special because it provides the clearest LRD analogue, combining the coolest enriched terminal state with a compact optically thick shell.

The chemical columns of Table~\ref{tab:A2} add a second strong conclusion. The Pop III shell is chemically distinct, with very large positive \(\log(C/O)\) and \(\log(N/O)\), marking the extreme primary-processing limit. Among the metal-enriched models, the shells remain H/He dominated, but the detailed CNO pattern changes with both metallicity and ejection depth. The \(Z=10^{-5}\,Z_\odot\) and \(Z=10^{-4}\,Z_\odot\) models show appreciable branch sensitivity, especially in \(\log(N/O)\), whereas the \(Z=10^{-3}\,Z_\odot\) and \(Z=10^{-2}\,Z_\odot\) cases give robust nitrogen-rich shells. For the \(Z=10^{-3}\,Z_\odot\) model, \(\log(N/O)=0.334\)–0.359. For the \(Z=10^{-2}\,Z_\odot\) model, the final-shell values are \(\log(N/O)\simeq0.13\) and \(0.10\) for the fiducial and upper branches, respectively, consistent with the main-text shell analysis. The He/H ratio varies much less than the CNO ratios, confirming that these events expel outer H/He envelope material rather than exposing a fully processed core. Among the metal-enriched models, Ne/O is also more stable than N/O, which makes N/O the sharper abundance discriminator.

Taken together, Table~\ref{tab:A2} shows that late pulsational shell ejection is a generic feature of our SMS models, not a peculiarity of the \(Z=10^{-2}\,Z_\odot\) sequence. Metallicity does not decide whether such shells occur at all; it regulates how they occur. Lower and intermediate metallicities can eject more total mass and often do so through more numerous episodes, while finite-metallicity models near \(10^{-3}\)–\(10^{-2}\,Z_\odot\) produce the clearest combination of a cool inflated terminal state, a compact optically thick shell, and an observationally useful nitrogen-rich composition.

\subsection{GR stability analysis}
\label{app:extended_methodsA10}
At each stellar-evolution timestep, we assess the approach to collapse with three complementary GR stability criteria: the full linear adiabatic GR analysis of \citet{Saio2024}, the post-Newtonian instability estimate of \citet{Haem2021A&A...647A..83H}, and the GR radial-stability method of \citet{nag22}. The first two define the main evolutionary instability point used in the stellar-evolution analysis, while the Nagele et al.\ criterion is also used to select the snapshot for the hydrodynamic follow-up. In practice, this criterion identifies a slightly earlier unstable model and avoids remapping a profile with an artificially steep outer pressure gradient.

For the direct GR radial-stability check, we solve the Chandrasekhar pulsation equation \citep{chandra64},
\begin{equation}
\begin{split}
&e^{-2a-b}\frac{d}{dr}\!\left[
\frac{e^{3a+b}\Gamma_1 P}{r^2}
\frac{d}{dr}\!\left(e^{-a}r^2\xi\right)
\right]
-\frac{4}{r}\frac{dP}{dr}\,\xi \\
&\quad
+ e^{-2a+2b}\omega^2(P+\rho c^2)\xi
-\frac{8\pi G}{c^4}e^{2b}P(P+\rho c^2)\xi \\
&\quad
-\frac{1}{P+\rho c^2}\left(\frac{dP}{dr}\right)^2 \xi
= 0 ,
\end{split}
\label{eq:gr_stability}
\end{equation}
which has previously been applied to numerical SMS models \citep{Haem2021A&A...647A..83H,nag22}. Here \(\rho=\rho_{\rm baryon}(1+\epsilon/c^2)\) is the relativistic density, \(a\) and \(b\) are metric coefficients, and \(\xi(r)e^{i\omega t}\) is the radial displacement. The solutions form a discrete mode sequence with \(\omega_n^2 < \omega_{n+1}^2\). Instability therefore occurs once the fundamental mode satisfies \(\omega_0^2<0\). We have cross-checked this criterion against the hydrodynamic collapse calculations described below.

\subsection{1D GR hydrodynamic follow-up}
\label{app:extended_methodsA11}
Once an unstable snapshot is selected with the GR stability analysis above, we remap the GENEC structure to a 1D Lagrangian GR hydrodynamics code \citep{nag20}. The code includes a 52-isotope nuclear network and thermal-neutrino cooling \citep{nag21}. We use this calculation as a consistency check on the linear stability analysis and to verify the final collapse outcome of the unstable SMS.

Figure~\ref{fig:gr_collapse} illustrates the collapse dynamics in the baseline run. The velocity profiles are initially close to homologous, with inward motion across most of the star. At later times, the inner regions accelerate more strongly than the outer layers, and the collapse becomes increasingly centrally concentrated as black-hole formation approaches. In the baseline non-rotating calculation, we find collapse rather than disruption.

\begin{figure}
\centering
\includegraphics[width=\columnwidth]{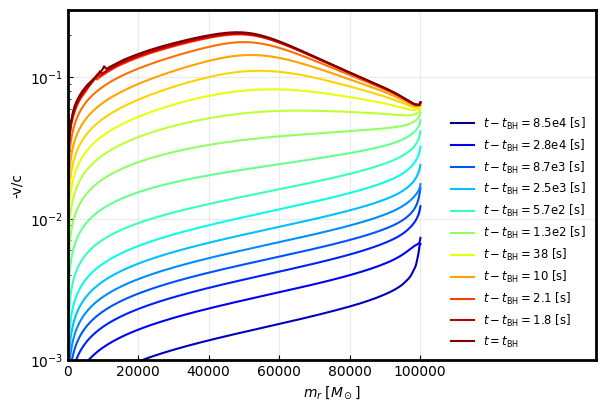}
\caption{Radial velocity profiles in units of $c$ as a function of enclosed mass coordinate for selected times in the 1D GR hydrodynamic follow-up of the unstable SMS model. The collapse is initially close to homologous, with inward motion across most of the star, and later becomes increasingly centrally concentrated as black-hole formation approaches.}
\label{fig:gr_collapse}
\end{figure}

\subsection{Why a truly cool hydrostatic photosphere is not required}

The shell-reprocessing picture above offers a natural explanation for the low apparent continuum temperatures inferred for some LRDs. It also avoids a difficulty faced by models in which the central source itself has a true hydrostatic photosphere at $T_{\rm eff}\sim3000$--$4000$~K.

A stellar photosphere in that temperature range should resemble a cool supergiant atmosphere. Such spectra are expected to show strong molecular absorption bands, especially from $TiO$, and often also from $C_2$ and $CN$ \citep{Levesque17}. These features should remain visible even at low spectral resolution. Even red supergiants with weak $CN$ still show clear $TiO$ structure \citep{2025ApJ...987..203G}. By contrast, a hotter source seen through an expanding dense shell can produce a much cooler continuum without requiring the underlying source itself to be a $3000$--$4000$~K star.

Dense shells also provide a natural route to reddening. A moderately reddened $T_{\rm eff}\sim7000$~K atmosphere can resemble a cooler atmosphere without strong molecular bands in low-resolution data. This helps explain why continuum-based temperature estimates alone may bias the source toward artificially low values. The low-redshift LRD analog $J1025+14$ supports this caution. Its continuum appears very cool, yet its absorption features imply substantially higher surface temperatures \citep{2025ApJ...980L..34L,2026MNRAS.545f2235J}.

{\setlength{\tabcolsep}{2.5pt}
\renewcommand{\arraystretch}{1.10}
\begin{deluxetable*}{ccccccccccccccc}
\tabletypesize{\fontsize{6.4}{7.4}\selectfont}
\label{tab:A2}
\tablecaption{Summary of the five $10^5\,M_\odot$ Phase XVI sequences as a function of metallicity. Listed are the number of fiducial ejection episodes, the total integrated ejected mass $\Delta M_{\rm ej}$, the largest single-episode ejecta mass $M_{\rm ep,max}$, the final stellar mass $M_{\rm f}$, the final effective temperature $\log T_{\rm eff,f}$, the final Eddington factor $\Gamma_{\rm Edd,f}$, the terminal escape speed $v_{\rm esc,last}$, the upper launch-speed proxy $v_{\rm launch}^{+}$, and the characteristic radius $R_{\rm sh}$ and electron-scattering optical depth $\tau_{\rm es}$ of the most relevant shell. The final columns give the integrated ejecta abundance ratios $\log(C/O)$, $\log(N/O)$, $\log(\mathrm{He}/\mathrm{H})$, and $\log(\mathrm{Ne}/\mathrm{O})$ as logarithmic number ratios. For cells with two entries, the upper and lower values correspond to the fiducial and upper-limit cases, respectively.}
\tablewidth{0pt}
\tablehead{\colhead{$Z/Z_\odot$} & \colhead{$N_{\rm ep}$} & \colhead{\shortstack[c]{$\Delta M_{\rm ej}$\\$[M_\odot]$}} & \colhead{\shortstack[c]{$M_{\rm ep,max}$\\$[M_\odot]$}} & \colhead{\shortstack[c]{$M_{\rm f}$\\$[M_\odot]$}} & \colhead{$\log T_{\rm eff,f}$} & \colhead{$\Gamma_{\rm Edd,f}$} & \colhead{\shortstack[c]{$v_{\rm esc,last}$\\$[\mathrm{km\,s^{-1}}]$}} & \colhead{\shortstack[c]{$v_{\rm launch}^{+}$\\$[\mathrm{km\,s^{-1}}]$}} & \colhead{\shortstack[c]{$R_{\rm sh}$\\$[\mathrm{pc}]$}} & \colhead{$\tau_{\rm es}$} & \colhead{$\log(C/O)$} & \colhead{$\log(N/O)$} & \colhead{$\log(\mathrm{He}/\mathrm{H})$} & \colhead{$\log(\mathrm{Ne}/\mathrm{O})$}}
\startdata
\rowcolor{smsZzero!24} $0$ & 14 & \shortstack[c]{2004.6\\2711.8} & \shortstack[c]{976.6\\1004.8} & 100662 & 4.872 & 0.977 & 7731 & 5058 & 1.4e-05 & $1.26\times10^{4}$ & \shortstack[c]{2.805\\2.797} & \shortstack[c]{3.542\\3.538} & \shortstack[c]{-1.083\\-1.083} & \shortstack[c]{-9.446\\-9.440} \\
\rowcolor{smsZ5!24} $10^{-5}$ & 24 & \shortstack[c]{6262.1\\14656.4} & \shortstack[c]{755.9\\1000.1} & 100888 & 4.018 & 1.005 & 1432 & 1291 & 1.2e-04 & $2.30\times10^{3}$ & \shortstack[c]{-0.756\\-0.832} & \shortstack[c]{-0.109\\0.145} & \shortstack[c]{-1.048\\-0.917} & \shortstack[c]{-0.542\\-0.528} \\
\rowcolor{smsZ4!24} $10^{-4}$ & 14 & \shortstack[c]{1854.8\\3336.2} & \shortstack[c]{990.5\\1000.9} & 100653 & 4.058 & 1.006 & 1574 & 1382 & 5.3e-05 & $2.25\times10^{3}$ & \shortstack[c]{-0.385\\-0.562} & \shortstack[c]{-0.519\\-0.253} & \shortstack[c]{-1.077\\-1.057} & \shortstack[c]{-0.545\\-0.535} \\
\rowcolor{smsZ3!24} $10^{-3}$ & 28 & \shortstack[c]{4827.4\\7196.6} & \shortstack[c]{998.9\\999.2} & 100185 & 4.066 & 0.994 & 1555 & 29656 & 9.3e-05 & 742.71 & \shortstack[c]{-0.249\\-0.246} & \shortstack[c]{0.334\\0.359} & \shortstack[c]{-0.659\\-0.630} & \shortstack[c]{-0.551\\-0.551} \\
\rowcolor{smsZ2!24} $10^{-2}$ & 4 & \shortstack[c]{479.9\\1039.6} & \shortstack[c]{348.0\\599.9} & 100254 & 3.965 & 0.980 & 1176 & 1358 & 1.5e-04 & $2.46\times10^{3}$ & \shortstack[c]{-0.232\\-0.263} & \shortstack[c]{0.13\\0.10} & \shortstack[c]{-0.966\\-0.969} & \shortstack[c]{-0.505\\-0.505} \\
\enddata
\tablecomments{For cells with two entries, the upper and lower values correspond to the fiducial and upper-limit cases, respectively. $R_{\rm sh}$ is in pc. Abundance ratios are logarithmic number ratios.}
\end{deluxetable*}
}

\end{document}